\documentclass[12pt,a4paper]{article}
\usepackage{graphicx}
\usepackage[dvips]{hyperref}

\newcommand{\be}{\begin{equation}}
\newcommand{\ee}{\end{equation}}
\newcommand{\ba}{\begin{eqnarray}}
\newcommand{\ea}{\end{eqnarray}}

\newcommand{\lla}{\langle\langle}
\newcommand{\rra}{\rangle\rangle}

\newcommand{\one}{\mathrm{I}}

\begin{document}

\begin{titlepage}
\begin{flushright}
LU TP 11-40\\
%arXiv:yymm.nnnn [hep-ph]\\
November 2011\\
\end{flushright}
\vfill
\begin{center}
{\Large\bf Two-Point Functions and S-Parameter\\[0.5cm]in QCD-like Theories}
\vfill
{\bf Johan Bijnens and Jie Lu}\\[0.3cm]
{Department of Astronomy and Theoretical Physics, Lund University,\\
S\"olvegatan 14A, SE 223-62 Lund, Sweden}
\end{center}
\vfill
\begin{abstract}
We calculated the vector, axial-vector, scalar and pseudo-scalar
two-point functions up to two-loop level 
in the low-energy effective field theory for three different
QCD-like theories. 
In addition we also calculated the pseudo-scalar decay constant $G_M$.
The QCD-like theories we used are those with fermions in
a complex, real or pseudo-real representation with in general n flavours.
These case correspond to global symmetry breaking pattern
of $SU(n)_L\times SU(n)_R\to SU(n)_V$, $SU(2n)\to SO(2n)$ or
$SU(2n)\to Sp(2n)$. We also estimated the S parameter for those
different theories.
\end{abstract}
\vfill
%\keywords{Technicolour and Composite Models, Spontaneous Symmetry Breaking,
%Lattice Quantum Field Theory, Chiral Lagrangians}
%
%{\bf PACS:}
%\vfill
\end{titlepage}

\tableofcontents

\section{Introduction}

The different global symmetry breaking patterns of QCD-like theories with
a vector-like gauge group
have been summarized in \cite{Dimopoulos,Peskin,Preskill} around 30 years ago.
The global symmetry and its spontaneous breaking
depend on whether the fermions live in a complex, real and pseudo-real
representation of the gauge group. For $n$ identical fermions
this corresponds to the symmetry breaking pattern
$SU(n)_L\times SU(n)_R\to SU(n)_V$, $SU(2n)\to SO(2n)$ and
$SU(2n)\to Sp(2n)$ respectively. 
These theories can be used to characterize some of technicolor models with
vector-like gauge bosons. 
QCD-like theories are also important in the
theory of finite baryon density.
Here the real and pseudo-real case allow to investigate the
mechanism of diquark condensate and finite density without the sign problem.
A main nonperturbative tool in studying strongly interacting theories
is lattice gauge theory. Numerical calculations are performed at finite
fermion mass and need in general to be extrapolated to the zero mass limit.
In the case of QCD Chiral Perturbation Theory (ChPT)
is used to help with this
extrapolation. Our work has the intention of providing similar formulas for
the QCD-like theories using the effective field theory (EFT) appropriate
for the alternative global symmetry patterns.

These EFT have been used at lowest order (LO) \cite{Kogut}
with earlier work to be found in \cite{Kogan,Leutwyler,SV} and some studies
at next-to-leading order (NLO) have also appeared \cite{GL2,GL4,Splittorff}.
The former two are the usual QCD case with $n$ flavours.
In our earlier papers \cite{paper1,paper2} we have systematically studied
the effective field theory of these three different QCD-like theories
to next-to-next-to-leading order (NNLO).
We managed to write the EFT of these cases in an extremely similar form.
We calculated the quark-antiquark condensates, the mass and decay constant
of the pseudo-Goldstone bosons \cite{paper1},
and meson-meson scattering \cite{paper2}. 
In this paper we extend the analysis
to two-point correlation functions. 
We obtain expressions for the
vector, axial-vector, scalar and pseudo-scalar two-point functions 
as well as the pion pseudo-scalar coupling $G_M$ to NNLO\footnote{We use
LO, NLO and NNLO as synomyms for order $p^2$, order $p^4$ and order $p^6$
calculations even if the order $p^2$ vanishes.} or order $p^6$.

In our earlier work \cite{paper1,paper2}, we called the three different cases
QCD or complex, adjoint or real and two-colour or pseudo-real.
In this paper we use only the latter, more general, terminology.

One motivation for this set of work was the study of strongly interacting
Higgs sectors, reviews are \cite{Techni1,Techni2}.
For any model beyond the Standard Model, 
passing the test of oblique corrections, or precision
LEP observables, is crucial \cite{Peskin:1991sw,Altarelli:1990zd}.
Over the years, the impact of the oblique corrections in 
those models have been studied quite intensively but in strongly
interacting cases mainly an analogy with QCD has been invoked.
Lattice gauge theory methods allow to study strongly interacting models
from first principles. The contributions from these theories to the
$S$-parameter can be calculated using the two-point functions studied
here and our formulas are useful to perform the extrapolation to
the massless case. This was in fact the major motivation for the present work
but we included the other two-point functions for completeness.

The paper is organized as follows. 
In Section \ref{sect:EFT} we give a brief introduction to EFT
for the three different cases.
Section \ref{sect:twopoint} is the main part
of the paper. We define the fermion currents and the two-point functions in
Section~\ref{sect:deftwopoint}.
In Sections \ref{sect:VV} to \ref{sect:PP}, we present 
the calculation of vector, axial-vector, scalar, pseudo-scalar two-point
functions. In Section \ref{sect:S}, we discuss the oblique corrections and
the S-parameter. Section \ref{sect:conclusions} summarizes our main results
and we present the definition.

\section{Effective Field Theory}
\label{sect:EFT}

In this section we briefly review the EFT of QCD-like
theories, the details can be found in the earlier paper \cite{paper1}.
The basic methods are those of Chiral Perturbation Theory \cite{Weinberg1,GL1}.
The counting of orders is in all cases the same as in ChPT, we count momenta
as order $p$ and the fermion mass $m$ as order $p^2$.

\subsection{Complex representation: QCD and CHPT}

The case of $n$ fermions in a complex representation is essentially like QCD.
The Lagrangian with external left and right vector, scalar and pseudos-calar
external sources, $l_\mu, r_\mu, s$ and $p$, is
\ba
\label{LagQCD}
\mathcal{L} &=&  \overline q_{Li}i\gamma^\mu D_\mu q_{Li} +
    \overline q_{Ri}i\gamma^\mu D_\mu q_{Ri}
   +\overline q_{Li}\gamma^\mu l_{\mu ij} q_{Lj}
   +\overline q_{Ri}\gamma^\mu r_{\mu ij} q_{Rj}
\nonumber\\&&
   -\overline q_{Ri}\mathcal{M}_{ij}q_{Lj}
   -\overline q_{Li}\mathcal{M}^\dagger_{ij}q_{Rj} 
    \qquad\qquad i,j = 1,2,...,n\,.
\ea
The covariant derivative is given by $D_\mu q = \partial_\mu q-i G_\mu q$,
and the mass matrix $\mathcal{M}=s-ip$. The sums shown are over the flavour
index. The sums over gauge group indices are implicit.

The Lagrangian (\ref{LagQCD}) has a symmetry $SU(n)_L\times SU(n)_R$
which is made local by the external sources \cite{GL2,GL1}.
The quark-anti-quark condensate $\langle \bar q q\rangle$
breaks $SU(n)_L\times SU(n)_R$ spontaneously to the diagonal
subgroup $SU(n)_V$.
According to the Nambu-Goldstone theorem, $n^2-1$ Goldstone Bosons will thus
be generated. We add a small fermion mass $m$ explicitly by setting
$s = m + s$. This mass term explicitly breaks the symmetry
$SU(n)_L\times SU(n)_R$ down to $SU(n)_V$ as well and gives the Goldstone
bosons a small mass.

The Goldstone boson manifold $SU(n)_L\times SU(n)_R/SU(n)_V$
can be parametrized by
\be
u = \exp\left(\frac{i}{\sqrt{2}\,F}\pi^a T^a\right)\qquad a = 1,2,...,n^2-1\,.
\ee
The $T^a$ are the generators of $SU(n)$ normalized to 
$\langle T^a T^b \rangle = \delta^{ab}$. The notation $\langle A \rangle$
stands for the trace over flavour indices.
$u$ transforms under $g_L\times g_R\in 
SU(n)_L\times SU(n)_R$ as $u\to g_R u h^\dagger = h u g_L^\dagger$ where
$h$ is the ``compensator'' and is a function of $u$, $g_L$ and $g_R$.
The methods are those of \cite{CCWZ}, but we use the notation of
\cite{BCE1,BCE2}. 
We can construct quantities which
transform under the unbroken group $H$ as : $O \to hOh^\dagger$
\ba
\label{QCDstandard}
u_\mu &=&i[u^\dagger(\partial_\mu-ir_\mu)u - u(\partial_\mu-l_\mu)u^\dagger]
\,,\nonumber\\
\nabla_\mu O &=&
\partial_\mu O+\Gamma_\mu O-O\Gamma_\mu
\nonumber\,,\\
\chi_\pm &=& u^\dagger \chi u^\dagger\pm u\chi^\dagger u
\,,\nonumber\\
f_{\pm\mu\nu}&=&
 u l_{\mu\nu} u^\dagger \pm u^\dagger r_{\mu\nu} u
\,.
\ea
The field strengths $l_{\mu\nu}$ and $r_{\mu\nu}$ are
\ba
l_{\mu\nu}& =& \partial_\mu l_\nu -  \partial_\nu l_\mu - i[l_\mu, l_\nu]\,,
\nonumber\\
r_{\mu\nu}& =& \partial_\mu r_\nu -  \partial_\nu r_\mu - i[r_\mu, r_\nu] \,.
\ea
The covariant derivative $\nabla_\mu$ contains
\be
\label{QCDGamma}
\Gamma_\mu =
\frac{1}{2}[u^\dagger(\partial_\mu-ir_\mu)u+u(\partial_\mu-l_\mu)u^\dagger]
\,.
\ee
$\chi$ contains the matrix $\mathcal{M}$,
which is the combination of scalar and pseudo-scalar sources
\be
\chi = 2B_0 \mathcal{M} = 2B_0(s-ip).
\ee
Using the quantities in (\ref{QCDstandard}), we can find the leading order,
$p^2$, Lagrangian
which is invariant under Lorentz and chiral symmetry:
\be
\label{QCDLO}
\mathcal{L}_2 = \frac{F^2}{4}\langle u_\mu u^\mu +\chi_+\rangle\,.
\ee
The subscript ``2'' stands for the order of $p^2$. The $p^4$ and $p^6$
Lagrangian will be explained
in Section \ref{sect:highorders}.

\subsection{Real and Pseudo-Real representation}

The case of $n$ fermions in a real or pseudo-real representation
of the gauge group we can treat in a similar way
as the complex case. 
In the real case, the global symmetry breaking pattern is $SU(2n)\to SO(2n)$,
and the number of generated Goldstone bosons is $2 n^2+n-1$.  
In the pseudo-real case, 
the symmetry breaking is $SU(2n)\to Sp(2n)$, and the number of
generated Goldstone is $2n^2-n-1$.
In both cases anti-fermions are in the same representation of the gauge group
and can be put together in a $2n$ vector $\hat q$, see \cite{paper1} 
for more details.

The condensate can now be a diquark condensate as well
as a quark-antiquark condensate. Our choice of vacuum corresponds to
a quark-anti-quark condensate. In terms of the $2n$ fermion vector $\hat q$
they can be written as
\begin{eqnarray}
\mathrm{Real}: &\langle{ \hat q^T C J_S \hat q}\rangle+ \mathrm{h.c.}&
\qquad
 J_S=
\left(\begin{array}{cc} 0 & \one\\
\one & 0\end{array}\right),
\\
\mathrm{Pseudo-Real}: &\langle \hat q_\alpha\epsilon_{\alpha\beta} C J_A \hat q_\beta\rangle+ \mathrm{h.c.}
&\qquad
J_A=
\left(\begin{array}{cc} 0 & -\one\\
\one & 0\end{array}\right).
\end{eqnarray}
Here $C$ is the charge conjugation operator.
$J_S$ and $J_A$ are symmetric and anti-symmetric
$2n\times2n$ matrices, $I$ is the $n\times n$ unit matrix. 
Since $J_S$ and $J_A$ often appear in the same
way in the expressions, we use $J$ for both cases unless
a distinction is needed.

The generators, $T^a$, of the global symmetry group $SU(2n)$ can be separated
into belonging to the broken, $X^a$, or unbroken part, $Q^a$.
They satisfy the following relations with $J$:
\be
\label{commutatorsAdjoint}
J Q^a = - Q^{aT} J\,,\qquad J X^a = X^{aT} J\,,
\ee
The Goldstone boson manifold can be parametrized with
\be
U = u J u^T\to g U g^T\ ,\qquad\mathrm{with}\qquad
u=\exp\left(\frac{i}{\sqrt{2}\,F}\pi^a X^a\right)\,.
\ee
where $J=J_S$ and $a$ runs from $1$ to $2n^2+n-1$ for the real case
and $J=J_A$ and $a$ runs from $1$ to $2n^2-n-1$ for the pseudo-real case.

In our earlier paper \cite{paper1}, we constructed quantities similar to
those in (\ref{QCDstandard}--\ref{QCDGamma})
\ba
\label{defumuadjoint}
u_\mu &=& 
i[u^\dagger(\partial_\mu-i V_\mu)u - u(\partial_\mu+iJ V^T_\mu J)u^\dagger]
\,,\nonumber\\
\Gamma_\mu &=& 
\frac{1}{2}[u^\dagger(\partial_\mu-i V_\mu)u 
            + u(\partial_\mu+iJV^T_\mu J)u^\dagger]
\nonumber\,. \\
f_{\pm\mu\nu} &=&  
J u V_{\mu\nu} u^\dagger J\pm  u V_{\mu\nu} u^\dagger
\,, \nonumber\\
\chi_\pm &=& u^\dagger\chi J u^{\dagger}  \pm  u J\chi^\dagger u
\,.
\ea
The $2n\times 2n$ matrix $V_\mu$ includes the left and right-handed
external sources
\be
 V_\mu =  \left(\begin{array}{cc} r_\mu & 0 \\
     0 & -l_\mu^T \end{array}\right)
\ee
and $V_{\mu\nu}$ is the field strength
\be
V_{\mu\nu}=\partial_\mu V_\nu-\partial_\nu V_\mu-i
\left(V_\mu V_\nu- V_\nu V_\mu\right)\, .
\ee
$\chi$ include the matrix $\hat\mathcal{M}$ via $\chi = 2B_0 \hat\mathcal{M}$
\cite{paper1}.
Those quantities behave similarly as those (\ref{QCDstandard}) if we take
\be
-J V^T_\mu J \to l_\mu  \,,\qquad V_\mu \to r_\mu  \,.
\ee
With this correspondence,
the Lagrangian of the real and pseudo-real case has the same form as
the complex one.
However one has to remember there are differences in the generators,
external sources, coupling constants,\ldots.
Anyway, now we can use the techniques of ChPT
to perform the calculations.

\subsection{High Order Lagrangians and Renormalization}
\label{sect:highorders}

Using Lorentz and chiral invariance,
we can write down the $p^4$ EFT lagrangian \cite{GL1} for all three cases
using the quantities listed in (\ref{QCDstandard}) and (\ref{defumuadjoint}):
\ba
\label{NLOlagrangian}
\mathcal{L}_4 &=&
L_0 \langle u^\mu u^\nu u_\mu u_\nu \rangle
+L_1 \langle u^\mu u_\mu\rangle\langle u^\nu u_\nu \rangle
+L_2 \langle u^\mu u^\nu\rangle\langle u_\mu u_\nu \rangle
+L_3 \langle u^\mu u_\mu u^\nu u_\nu \rangle
\nonumber\\&&
+L_4  \langle u^\mu u_\mu\rangle\langle\chi_+\rangle
+L_5  \langle u^\mu u_\mu\chi_+\rangle
+L_6 \langle\chi_+\rangle^2
+L_7 \langle\chi_-\rangle^2
+\frac{1}{2} L_8 \langle\chi_+^2+\chi_-^2\rangle
\nonumber\\&&
-i L_9\langle f_{+\mu\nu}u^\mu u^\nu\rangle
+\frac{1}{4}L_{10}\langle f_+^2-f_-^2\rangle
+H_1\langle l_{\mu\nu}l^{\mu\nu}+r_{\mu\nu}r^{\mu\nu}\rangle
+H_2\langle\chi\chi^\dagger\rangle\,.
\ea
To do the renormalization, we use the 
ChPT $\overline {\mathrm{MS}}$ scheme with dimensional regularization
\cite{GL1,GL2,BCE2}.
The bare coupling constants $L_i$ are defined as
\be
\label{defLir}
L_i = \left(c \mu\right)^{d-4}
\left[\Gamma_i\Lambda+L_i^r(\mu)\right] \, ,
\ee
where the dimension $d=4-2\epsilon$, and
\ba
\Lambda &=& {1\over16\pi^2(d-4)}\ ,\\
\ln c &=& -{1\over2}[\ln 4\pi+\Gamma^\prime(1)+1]\ .
\ea
The coefficients $\Gamma_i$ for the complex case have been obtained in
\cite{GL2}, for the real and pseudo-real case we have calculated them
earlier in \cite{paper1}. However, there are mistakes
in the coefficients of $L_9, L_{10}$ and $H_1$  in the Table 1 of \cite{paper1}.
These mistakes had no effects on our previous calculations.
We therefore list all the coefficients here again in Table \ref{tabdivergence}.
\begin{table}
\begin{center}
\begin{tabular}{|c|c|c|c|}
\hline
i & complex & real & pseudo-real \\
\hline
0  & $n/48$              & $(n+4)/48$         & $(n-4)/48$         \\
1  & $1/16$              & $1/32$             & $1/32$             \\
2  & $1/8$               & $1/16$             & $1/16$             \\
3  & $n/24$              & $(n-2)/24$         & $(n+2)/24$         \\
4  & $1/8$               & $1/16$             & $1/16$             \\
5  & $n/8$               & $n/8$              & $n/8$              \\
6  & $(n^2+2)/(16 n^2)$  & $(n^2+1)/(32 n^2)$  & $(n^2+1)/(32 n^2)$ \\
7  & 0                   & 0                  & 0                  \\
8  & $(n^2-4)/(16n)$     & $(n^2+n-2)/(16n)$   & $(n^2-n-2)/(16n)$  \\
9  & $n/12$              & $(n+1)/12$          & $(n-1)/12$        \\
10 & $-n/12$             & $-(n+1)/12$         & $-(n-1)/12$       \\
1' & $-n/24$             & $-(n+1)/24$         & $-(n-1)/24$       \\
2' & $(n^2-4)/(8n)$      & $(n^2+n-2)/(8n)$    & $(n^2-n-2)/(8n)$   \\
\hline
\end{tabular}
\end{center}
\caption{\label{tabdivergence} The coefficients $\Gamma_i$
for the three cases that are needed to absorb the divergences at NLO.
The last two lines correspond to the terms with $H_1$ and $H_2$.
This is the same as Table 1 in \cite{paper1} but with the
error for $L_9, L_{10}$ and $H_1$ corrected.}
\end{table}

The $p^6$ Lagrangian for the complex case and general $n$
has been obtained in \cite{BCE1},
it contains 112+3 terms. 
The divergence structure of the bare coupling constants $K_i$ in the $p^6$
can be written as\footnote{The $K_i$ have been made dimensionless by
including a factor of $1/F^2$ explicitly in the order $p^6$
Lagrangian.}
\ba
\label{defKir}
K_i = \left(c\mu\right)^{2(d-4)}
\left[K^r_i-\Gamma_i^{(2)}\Lambda^2
-\left(\frac{1}{16\pi^2}\Gamma_i^{(1)}+\Gamma_i^{(L)}
\right)\Lambda
\right]\,.
\ea
The coefficients $\Gamma_i^{(2)}$, $\Gamma_i^{(1)}$ and
$ \Gamma_i^{(L)}$ for the complex case have been obtained in \cite{BCE2}.

For the real and pseudo-real case, the $p^6$ Lagrangian has the same form 
as in the complex case but some terms might be redundant.
The divergence structure as given in (\ref{defKir})
still holds but the coefficients are not known.
One check on our results that remains is that all the non-local divergences
cancel.

\section{Two-Point Functions}
\label{sect:twopoint}

\subsection{Definition}
\label{sect:deftwopoint}

The effective action of the fermion level theory with external sources is
\be
\exp\{iZ(l_\mu,r_\mu,s,p)\} = \int {\cal D}q {\cal D}\bar{q} {\cal D}G_\mu
\,\exp\bigg\{i\int d^4x {\cal L}_{QCD}(q,\bar q,G_\mu,l_\mu,r_\mu,s,p) \bigg\}
\ee
At low energies, i.e. below 1 GeV in QCD,
the effective action can be obtained also from the low-energy effective theory
\be
\exp\{iZ(l_\mu,r_\mu,s,p)\} = \int {\cal D}U \,
\exp\bigg\{i\int d^4x {\cal L}_{eff}(U,l_\mu,r_\mu,s,p)\bigg\}\,. 
\ee
With this effective action, the n-point Green functions can be easily
derived by taking the
functional derivative w.r.t. the external sources of $Z(J)$
\be
G^{(n)}(x_1,\ldots,x_n) =
\frac{\delta^n}{i^n\delta j(x_1) \ldots \delta j(x_n)} Z[J]
\bigg\vert_{J=0}\,.
\ee
Here $j$ stands for any of the external sources $l_\mu,r_\mu,s,p$
and $J$ for the whole set of them.

The vector current $v_\mu$ and axial-vector current $a_\mu$ are included via
\be
l_\mu = v_\mu-a_\mu\,,\qquad r_\mu = v_\mu+a_\mu\,.
\ee

In this paper we will calculate the two-point functions of vector,
axial-vector, scalar and pseudo-scalar currents.
The fermion currents in the complex case are defined as
\ba
%\label{defcurrents}
\label{vectorcurrent}
V_\mu^{a}(x) &=& \overline{q}_i T^a_{ij} \gamma_\mu q_j\;, \\
\label{axialvectorcurrent}
A_\mu^{a}(x) &=& \overline{q}_i  T^a_{ij} \gamma_\mu  \gamma_5 q_j\;, \\
\label{scalarcurrent}
S^{a}(x) &=& -\overline{q}_i  T^a_{ij}q_j\;, \\
\label{pseudoscalarcurrent}
P^{a}(x) &=& i\overline{q}_i T^a_{ij} \gamma_5 q_j\;.
\ea
$T^a$ is an $SU(n)$ generator\footnote{We have defined here the current
with fermion-anti-fermion operators, hence the $SU(n)$ for $n$ fermions.
For the real and pseudo-real case, the unbroken symmetry relates
them also to difermion ot dianti-fermmion operators.}
or in addition for the singlet
scalar and pseudo-scalar current the unit matrix which we label by $T^0$.
These currents also exist the for real and pseudo-real case.
In this case also currents with two fermions or two anti-fermions exist.
These can be combined with those above. The generators can then become
$SU(2n)$ generators. All conserved generators are like the vector or scalar case
while the broken generators are like the axial-vecor or pseudo-scalar case.
All those cases are related to the ones with the currents of 
(\ref{vectorcurrent})-(\ref{pseudoscalarcurrent}) via transformations
under the unbroken part of the symmetry group.

The definitions of the two-point functions are
\ba
\label{deftwop}
\Pi_{Va\mu\nu}(q) &\equiv&
i\int d^4x\; e^{iq\cdot x}\;\langle 0|T(V_\mu^a(x)V_\nu^a(0))^\dagger|0\rangle
\,,\nonumber\\
\Pi_{Aa\mu\nu}(q) &\equiv&
i\int d^4x\; e^{iq\cdot x}\;\langle 0|T(A_\mu^a(x)A_\nu^a(0))^\dagger|0\rangle
\,,\nonumber\\
\Pi_{SMa\mu}(q) &\equiv&
i\int d^4x\; e^{iq\cdot x}\;\langle 0|T(V^a_\mu(x)S^a(0))^\dagger|0\rangle
\,,\nonumber\\
\Pi_{PMa\mu}(q) &\equiv&
i\int d^4x\; e^{iq\cdot x}\;\langle 0|T(A^a_\mu(x)P^a(0))^\dagger|0\rangle
\,,\nonumber\\
\Pi_{Sa}(q) &\equiv&
i\int d^4x\; e^{iq\cdot x}\;\langle 0|T(S^a(x)S^a(0))^\dagger|0\rangle
\,,\nonumber\\
\Pi_{Pa}(q) &\equiv&
i\int d^4x\; e^{iq\cdot x}\;\langle 0|T(P^a(x)P^a(0))^\dagger|0\rangle
\,.
\ea
Using Lorentz invariance the two-point functions with vectors and axial-vectors
can be decomposed in scalar functions
\be
\Pi_{Va\mu\nu} = (q_\mu q_\nu-q^2 g_{\mu\nu})\Pi_{Va}^{(1)}(q^2)
 +q_\mu q_\nu \Pi_{Va}^{(0)}(q^2)\;.
\ee
where $\Pi_{Va}^{(1)}(q^2)$ is the transverse part and
$\Pi_{Va}^{(0)}(q^2)$ is the longitudinal part or alternatively the spin one and
spin 0 part.
 The same definition holds
for the axial-vector two-point functions.
The mixed functions can be decomposed as
\ba
\Pi_{SMa\mu} &=& q_\mu \Pi_{SMa}
\,,\nonumber\\
\Pi_{PMa\mu} &=& iq_\mu \Pi_{PMa}\,.
\ea
Using the divergence of fermion currents and equal time commutation relations,
we find that some two-point functions are related to each other by
Ward identities.
In the equal mass case considered here, they are
\ba
\label{Wardi}
\Pi_{Va}^{(0)} &=&\Pi_{SMa}=0
\,,\nonumber\\
q^2 \Pi_{Aa}^{(0)} &=& 2m\Pi_{PMa}
\,,\nonumber\\
q^4\Pi^{(0)}_{Aa} &=& 4m^2 \Pi_{Pa} + 4m \langle \bar q q \rangle \,.
\ea
The vacuum expectation value is the single quark-anti-quark one.
We will use the last relation to double check our results of
axial-vector and pseudo-scalar two-point functions.

The mixed two-point functions, $\Pi_{SMa}$ and $\Pi_{PMa}$
we do not discuss further since they are fully given by the Ward identities.

\subsection{The Vector Two-Point Function}
\label{sect:VV}

The vector two-point function is defined in (\ref{deftwop}).
The longitudinal part vanishes for all three cases because of the Ward
identities.

\begin{figure}
 \begin{center}
\includegraphics[width=10cm]{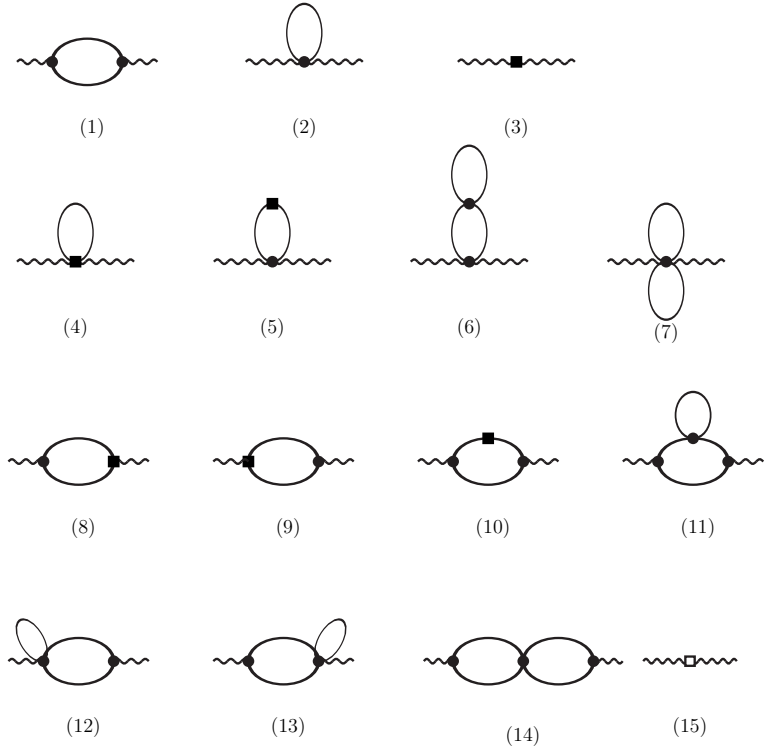}
\caption{The diagrams for the vector two-point function.
A filled circle is a vertex
from ${\cal L}_{2}$, a filled square  a vertex from ${\cal
L}_{4}$, and an open square a vertex from ${\cal L}_6$.
The top line is order $p^4$. The remaining ones are order $p^6$.}
\label{vv2point}
\end{center}
\end{figure}
The Feynman diagrams for the vector two-point function are
listed in Fig.~\ref{vv2point}. There is no diagram at lowest order.
The diagrams at NLO are (1--3) in Fig.~\ref{vv2point}.
The NNLO diagrams are (4--15).
The 3-flavour QCD case is known to NNLO \cite{ABT1,GK1}.

We have rewritten the results in terms of the physical mass and decay constant.
For these we use the notation $M_M$ and $F_M$ rather than the $M_\mathrm{phys}$
and $F_\mathrm{phys}$ used in \cite{paper1,paper2}. 
Their expression in terms of the lowest order quantities $F$
and $M^2=2B_0 m$ can be found in \cite{paper1}.
We also use
the quantities
\be
L = \frac{1}{16\pi^2}\log\frac{M_M^2}{\mu^2}\quad\mathrm{and}\quad
\pi_{16} = \frac{1}{16\pi^2}\,.
\ee
The loop integral $\overline B_{22}$ is defined in Appendix \ref{oneloopint}.

The results up to NNLO for three different cases are listed below,
where the first line in each case
is the NLO contributions, the rest are NNLO contributions.
\ba
&&\mathbf{Complex}
\nonumber\\
\quad \Pi^{(1)}_{VV}&=& -{n\over q^2}
\left[4   \overline B_{22}(M^2_M,M^2_M,q^2)+ 2 L M^2_M \right]
                   -4 L^r_{10}-8 H^r_1
\nonumber\\
&&+{1\over F^2_M}\Bigg\{
\left(\frac{4 M^2_M}{ q^2} L n^2- 16 L^r_{9} n \right)
   \overline B_{22}(M^2_M,M^2_M,q^2)
+\frac{4 n^2 }{ q^2}[\overline B_{22}(M^2_M,M^2_M,q^2)]^2
\nonumber\\
&&\quad+ {M^4_M\over q^2}L^2  n^2-   8q^2 K^r_{115}
+  8M^2_M(  L L^r_{10} n-4 K^r_{81}-4 K^r_{82} n)
 \Bigg\}
\,,\\
 %%%%%%%%%%%%%%%%%
&&\mathbf{Real}
\nonumber\\
\Pi^{(1)}_{VV}&=&
-\frac{1}{q^2}(n+1)\left[4\overline B_{22}(M^2_M,M^2_M,q^2)+ 2M^2_M L \right]
-4 L^r_{10}-8 H^r_1  
\nonumber\\
&&+{1\over F^2_M}\Bigg\{\left[
\frac{ 4M^2_M}{ q^2}L (n+1)^2 -16 (n +1)L^r_9 \right] \overline B_{22}(M^2_M,M^2_M,q^2) 
\nonumber\\
&&\quad+\frac{4}{ q^2}(n+1)^2[\overline B_{22}(M^2_M,M^2_M,q^2)]^2
+\frac{ M^4_M}{ q^2}L^2 (n+1)^2
\nonumber\\
&&\quad
- 8 q^2 K^r_{115}
+8M^2_M [L L^r_{10}(n+1)-4 K^r_{81}-8 K^r_{82} n]
\Bigg\}\,,
\\
%%%%%%%%%%%%%%%%%%%
&&\mathbf{Pseudo-Real}
\nonumber\\
\Pi^{(1)}_{VV}&=&-\frac{1}{q^2} (n-1)\left[
4 \overline B_{22}(M^2_M,M^2_M,q^2)+  2M^2_M L \right]
-4 L^r_{10}-8 H^r_1
\nonumber \\
&&+{1\over F^2_M}\Bigg\{
\left[
\frac{4M^2_M}{ q^2} L  (n-1)^2
- 16 L^r_9( n-1)
\right]\overline B_{22}(M^2_M,M^2_M,q^2)
\nonumber\\
&&
\quad+\frac{4}{ q^2}(n-1)^2[\overline B_{22}(M^2_M,M^2_M,q^2)]^2
+\frac{ M^4_M}{ q^2}L^2 (n-1)^2
\nonumber\\
&&\quad
- 8 q^2 K^r_{115}
+8M^2_M [L L^r_{10}(n-1)-4 K^r_{81}-8 K^r_{82} n]\,.
\ea
The complex result with $n=3$ agrees with \cite{ABT1} when the masses
there are set equal.

\subsection{The Axial-Vector Two-Point Function}
\label{sect:AA}

The axial-vector two-point function is defined in (\ref{deftwop}).
Similar to the vector two-point function, it also can be decomposed in
a transverse and longitudinal part.
\be
\Pi^{\mu\nu}_{AA} = (q^\mu q^\nu-q^2 g^{\mu\nu})\Pi_{AA}^{(1)}(q^2)
 +q^\mu q^\nu \Pi_{AA}^{(0)}(q^2)\;.
\ee

\begin{figure}
\begin{center}
\includegraphics[width=11cm]{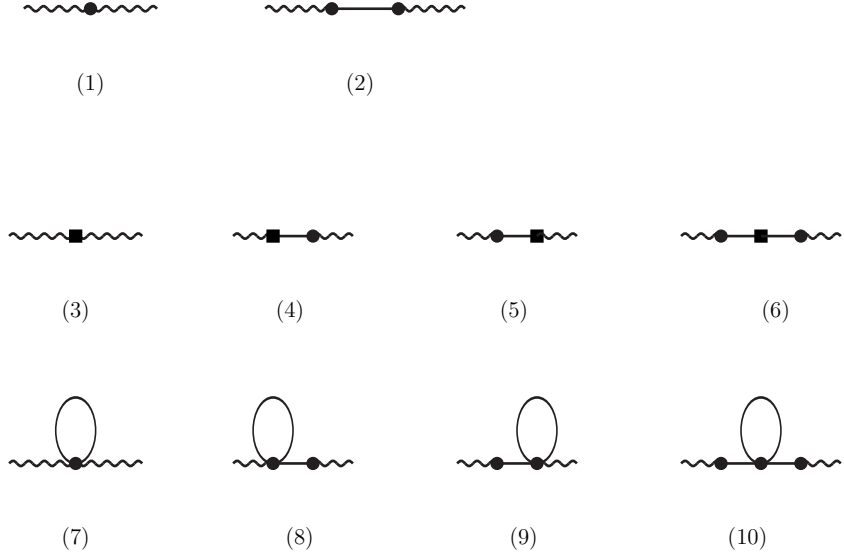}
\end{center}
\caption{The axial-vector two-point function at LO and NLO.
The filled circle is a vertex from ${\cal L}_{2}$, The
filled square is a vertex from ${\cal L}_{4}$, and the open square
is a vertex from ${\cal L}_6$.}.
\label{aa2pointp2p4}
\end{figure}
The diagrams contributing at LO are shown in (1--2) in Fig.~\ref{aa2pointp2p4}.
The LO results are the same for all three cases. The result is
\ba
\label{AAp2}
\Pi^{\mu\nu}_{AA}(q^2) =
2 F\left(g^{\mu\nu} -q^\mu q^\nu  {1\over q^2 - M^2} \right) \;.
\ea
$F$ and $M$ are the LO decay constant and mass respectively.
Note that in the massless limit this has only a transverse part as follows
from the Ward identities.

\begin{figure}
\begin{center}
\includegraphics[width=0.95\textwidth]{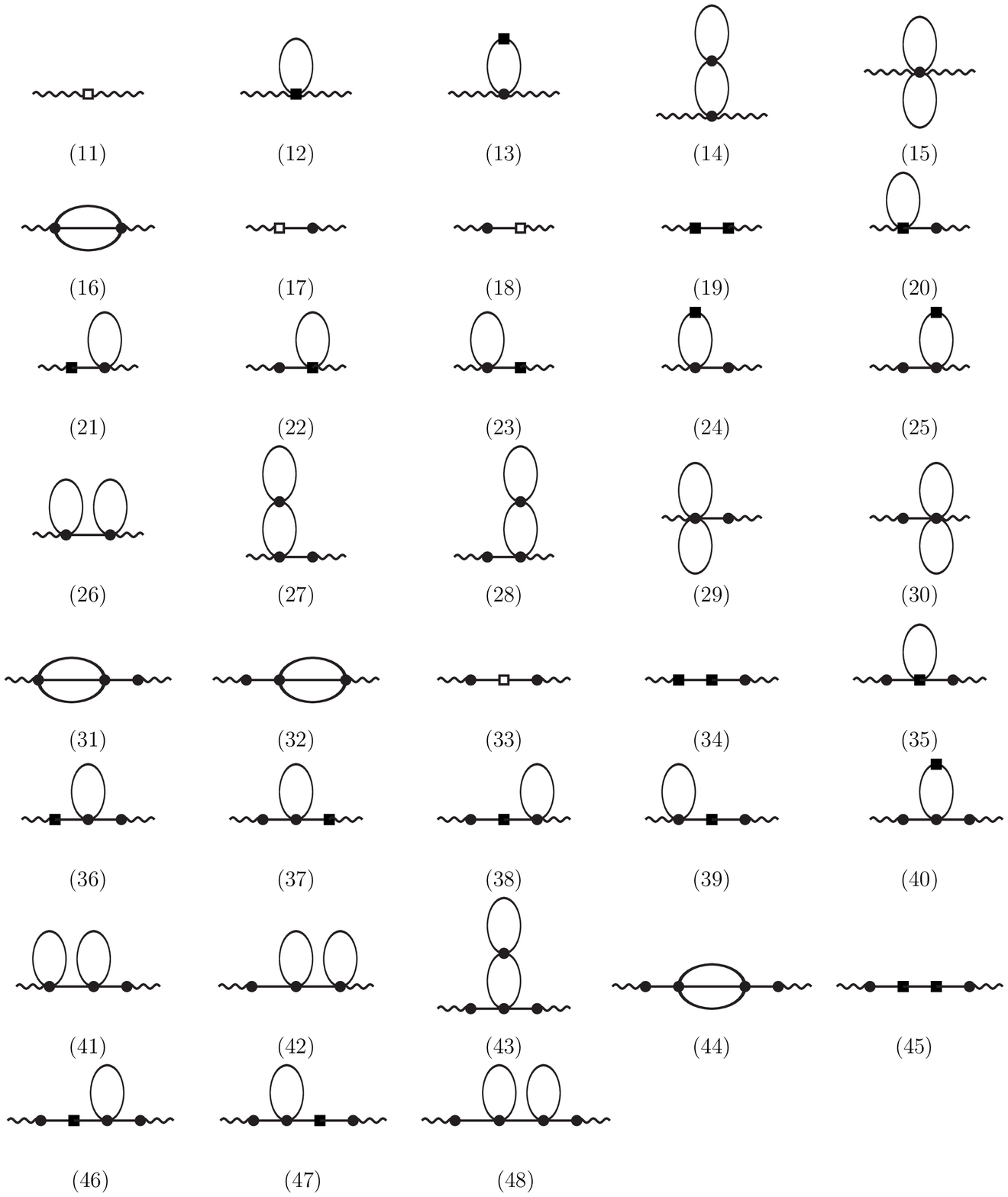}
\end{center}
\caption{\label{aa2pointp6}
The axial-vector two-point function at NNLO.
The filled circle is a vertex from ${\cal L}_{2}$, the filled square is
a vertex from ${\cal L}_{4}$, and the open square is a vertex from
${\cal L}_6$}
\end{figure}

The diagrams at NLO are (3--10) in Fig.~\ref{aa2pointp2p4}
and the NNLO diagrams are
(11--48) in Fig.~\ref{aa2pointp6}.

Many of the diagrams are one-particle-reducible and at first sight have
double and triple poles at $q^2=M^2$. From general properties of field theory
these should be resummable in into a single pole at the physical mass, 
$q^2=M_M^2$ and a nonsingular part that only has cuts. The residue at
the pole is the decay constant squared. We must thus find contributions
that allow for the last term in (\ref{AAp2}) the lowest order $F^2,M^2$
to be replaced by $F_M^2,M^2_M$. It turns out to be advantageous to also
do this in the first term. Most of the corrections are already included
in this way.

At NLO the remaining part is only from the tree level diagram (3) in
Fig.~\ref{aa2pointp2p4} and is
\ba
\Pi^{(1)}_{AA} =4 L^r_{10}-8 H^r_1 \;,
\qquad
\Pi^{(0)}_{AA}= 0     \;.
\ea
So we can express our result up to NNLO as
\ba
\label{AAp6}
\Pi^{\mu\nu}_{AA}(q^2) &=&
2 F_M^2\left(g^{\mu\nu} -q^\mu q^\nu  {1\over q^2 - M^2_M} \right)
 + (q^\mu q^\nu-q^2 g^{\mu\nu})(4L^r_{10}-8 H^r_1)
\nonumber\\
&&+{1\over F^2_M }\Bigg[(q^\mu q^\nu-q^2 g^{\mu\nu})\hat \Pi_{AA}^{(1)}(q^2)
 +q^\mu q^\nu \hat \Pi_{AA}^{(0)}(q^2) \Bigg]\;.
\ea
The $\hat\Pi_{AA}^{(0)}(q^2)$ and $\hat\Pi_{AA}^{(1)}(q^2)$ are the remainders
at NNLO and have no singularity at $q^2=M^2_M$.

The transverse part can be obtained from the part
containing $g_{\mu\nu}$ as an overall factor. So the transverse part
cannot come from the one-particle reducible diagrams and only gets contributions
from diagrams (11--16) at NNLO. The sunset integrals 
$H^F$ and $H^F_{21}$ appearing in the
results are defined in Appendix~\ref{sunsetint}.

The longitudinal part gets at NNLO contributions from
all diagrams shown in Fig~\ref{aa2pointp6}. In order to rewrite the results
into the single pole we need to expand the integrals around the mass.
This introduces derivatives of the sunsetintegrals. These always show
up in the combinations $H^M$ and $H^M_{21}$ defined in Appendix~\ref{sunsetint}.

\ba
&&\mathbf{Complex}\nonumber\\
\hat\Pi^{(1)}_{AA}(q^2)&=&
\frac{n^2}{2}\left[
\frac{M^2_M}{q^2} H^F(M^2_M,M^2_M,M^2_M,q^2)
- H^F_{21}(M^2_M,M^2_M,M^2_M,q^2)\right]
\nonumber\\
&&+M^2_M\left[\frac{L^2 n^2}{6}-8n L L^r_{10}
- 32 \left(K^r_{102}+K^r_{103} n+K^r_{17}+K^r_{18} n \right)\right]
\nonumber\\
&&
- q^2 (16 K^r_{109}+8 K^r_{115})
-\frac{M^4_M}{ q^2} \left(8 K^r_{113}+\frac{3}{2}  n^2L^2\right)
\nonumber\\
&&
+\pi_{16}L\left[\frac{M^4_M}{ q^2}   \left(\frac{13  n^2}{8}+\frac{2}{n^2}-\frac{2}{3}\right)
+\frac{ M^2_M}{6 } n^2 \right]
\nonumber\\
&&
+ \pi_{16}^2\Bigg[\frac{M^4_M}{ q^2}
\left(\frac{7}{2 n^2}-\frac{1}{8} n^2 \pi ^2-\frac{17 n^2}{64}-\frac{7}{6}\right)
\nonumber\\&&\quad
+ M^2_M   n^2 \left(\frac{  \pi ^2}{36}+\frac{1}{72}\right)
+\frac{n^2}{96 }q^2\Bigg]\,,
\\
\hat\Pi^{(0)}_{AA}(q^2)&=&
\left[ M^4_M \left(\frac{4}{3}-\frac{4}{n^2}\right)
-\frac{M^6_M}{2q^2} n^2\right] H^M(M^2_M,M^2_M,M^2_M,q^2)
\nonumber\\
&&
-\frac{3 M^4_M}{2 } n^2 H^M_{21}(M^2_M,M^2_M,M^2_M,q^2)+\frac{8M^4_M}{ q^2} K^r_{113} 
\nonumber\\
&&
+\frac{M^4_M}{ q^2}\left[ \pi_{16}L \left(\frac{2 }{3}-\frac{ n^2}{8}-\frac{2 }{n^2}\right)
+ \pi_{16}^2\left(\frac{7}{6}-\frac{15 n^2}{64}-\frac{7}{2 n^2}\right)\right]\,,
\\
%%%%%%%%%%%%%%%%%%%%%%%%
&&\mathbf{Real}\nonumber\\
\hat\Pi^{(1)}_{AA}(q^2)&=&
\frac{n}{2} (n+1)\left[
 \frac{M^2_M}{  q^2} H^F(M^2_M,M^2_M,M^2_M,q^2) 
 - H^F_{21}(M^2_M,M^2_M,M^2_M,q^2)\right]
\nonumber\\
&&  +  M^2_M\left[\frac{L^2}{6} n(n+1)- 8 nL L^r_{10}
- 32 (  K^r_{102}+ 2n K^r_{103}  +   K^r_{17}+ 2n K^r_{18} )\right] 
\nonumber\\
&&- q^2 (16 K^r_{109}+8 K^r_{115})
-\frac{ M^4_M}{  q^2}\left[ 8K^r_{113} +  \frac{3 }{2} n(n+1)L^2\right]
\nonumber\\
&&+\pi_{16}L\left[\frac{ M^4_M}{q^2} \left(\frac{13 n^2}{8}+\frac{1}{2 n^2}+\frac{43 n}{24}-\frac{1}{2 n}-\frac{1}{6}\right)
+   \frac{M^2_M}{6}n(n+1) \right] 
\nonumber\\
&&+ \pi_{16}^2 \Bigg[\frac{M^4_M}{ q^2}\left(
-\frac{1}{8} n^2 \pi ^2-\frac{n \pi ^2}{8}-\frac{17 n^2}{64}+\frac{7}{8 n^2}+\frac{5 n}{192}-\frac{7}{8 n}-\frac{7}{24}\right)
\nonumber\\
&&\quad + M^2_M  n(n+1) \left( \frac{  \pi ^2}{36} +\frac{1}{72} \right)
+ \frac{n}{96}(n+1)  q^2 \Bigg]\,,
\\
\hat\Pi^{(0)}_{AA}(q^2)&=&
 \left[-\frac{M^6_M}{  q^2} \frac{n}{2}(n+1)
+ M^4_M (n-1) \left(\frac{1}{n^2} - \frac{1}{3}\right)\right]
H^M(M^2_M,M^2_M,M^2_M,q^2)
\nonumber\\
&& -\frac{3}{2}  M^4_M  n(n+1) H^M_{21}(M^2_M,M^2_M,M^2_M,q^2)+\frac{8  M^4_M}{  q^2}K^r_{113}
\nonumber\\
&&+ \pi_{16}\frac{ M^4_M }{  q^2}L\left(-\frac{n^2}{8}-\frac{1}{2 n^2}-\frac{7 n}{24}+\frac{1}{2 n}+\frac{1}{6}\right)
\nonumber\\
&&+\pi_{16}^2\frac{M^4_M}{q^2} \left(-\frac{15 n^2}{64}-\frac{7}{8 n^2}-\frac{101 n}{192}+\frac{7}{8 n}+\frac{7}{24}\right)\,,\\
%%%%%%%%%%%%%%%%%%%%%%%%
&&\mathbf{Pseudo-Real}\nonumber\\
\hat\Pi^{(1)}_{AA}(q^2)&=&
\frac{n}{2} (n-1)\left[\frac{M^2_M}{ q^2}  H^F(M^2_M,M^2_M,M^2_M,q^2)
 - H^F_{21}(M^2_M,M^2_M,M^2_M,q^2)\right] 
\nonumber\\
&&- 32M^2_M ( K^r_{102}+2n K^r_{103}+  K^r_{17}+2nK^r_{18})
- 8q^2 ( 2K^r_{109}+ K^r_{115})
\nonumber\\
&&
 -\frac{8  M^4_M}{ q^2}K^r_{113}
-\frac{ M^4_M}{ q^2}\frac{3 n}{2}(n-1)L^2+   \frac{M^2_M}{6}n(n-1)L^2
- 8M^2_M nL L^r_{10}
\nonumber    \\
&&+ \pi_{16}\frac{L M^4_M}{ q^2} \left(\frac{13 n^2}{8}+\frac{1}{2 n^2}-\frac{43 n}{24}+\frac{1}{2 n}-\frac{1}{6}\right)+\pi_{16} L M^2_M \left(\frac{n^2}{6}-\frac{n}{6}\right) 
\nonumber\\
&&+\pi_{16}^2\Bigg[\frac{M^4_M}{ q^2}  \left(-\frac{1}{8} n^2 \pi ^2-\frac{17 n^2}{64}+\frac{7}{8 n^2}+\frac{n \pi ^2}{8}-\frac{5 n}{192}+\frac{7}{8 n}-\frac{7}{24}\right)\nonumber\\
&&\qquad+ M^2_M n(n-1)  \left(\frac{ \pi ^2}{36}+\frac{1}{72}\right)
+ q^2\frac{n}{96}(n-1) \Bigg]\,,
\\
\hat\Pi^{(0)}_{AA}(q^2)&=&
 \left[-\frac{M^6_M}{  q^2} \frac{n}{2}(n-1)
- M^4_M (n+1) \left(\frac{1}{n^2} - \frac{1}{3}\right)\right]
H^M(M^2_M,M^2_M,M^2_M,q^2)
\nonumber\\
&& -\frac{3}{2}  M^4_M  n(n-1) H^M_{21}(M^2_M,M^2_M,M^2_M,q^2)
+\frac{8  M^4_M}{  q^2}K^r_{113}
\nonumber\\
&&+ \pi_{16}\frac{ M^4_M }{  q^2}L\left(-\frac{n^2}{8}-\frac{1}{2 n^2}+\frac{7 n}{24}-\frac{1}{2 n}+\frac{1}{6}\right)
\nonumber\\
&&+\pi_{16}^2\frac{M^4_M}{q^2} \left(-\frac{15 n^2}{64}-\frac{7}{8 n^2}+\frac{101 n}{192}-\frac{7}{8 n}+\frac{7}{24}\right)\,.
\ea

The axial two-point function is known in 3-flavour ChPT \cite{ABT1,GK}. We have
checked that our result agrees with the one in \cite{ABT1} in the limit of equal
masses.

\subsection{The Scalar Two-Point Functions}
\label{sect:SS}

The scalar two-point function is defined in (\ref{deftwop}), which contains
the unbroken generator case ($T^a = Q^a$) and the singlet case ($a=0$).
 
The Feynman diagrams for both cases are 
the same as those for the vector two-point function shown in 
Figure~\ref{vv2point} except that diagrams (2) and (5--7) are absent.
Diagrams (1) and (3) are at NLO, and the diagrams (4) and (8--11) are at NNLO.

\subsubsection{$Q^a$ case}

The scalar two-point functions are similar to the vector two-point functions,
the LO results are zero for all the three cases since
the vertex at LO is absent.
We have rewritten again everything in terms of the physical mass and decay
constant, $M^2_M$ and $F_M$.
The results for the three cases are given below. The first line is
the NLO contribution and the remainder is the NNLO contribution.\\
\ba
&&\mathbf{Complex}\nonumber\\
\Pi_{SS}&=&B^2_0\left\{  8 H^r_2+16 L^r_{8} +\frac{1}{n}\left(n^2-4\right)
 \overline B(m^2,q^2)\right\}
\nonumber\\
&&+{B^2_0\over F^2_M }\Bigg\{
q^2 \left(8 K^r_{113}+32 K^r_{47}\right)
+ M^2_M \Bigg(192 K^r_{25}+64 K^r_{26} n\Bigg)
\nonumber\\&&
\quad
+ M^2_M L \Bigg[\left(\frac{64}{n}-32n\right) L^r_{8}
-64 L^r_{7}
+\left(n^2-4\right)\frac{16}{n} L^r_{5}\Bigg]
\nonumber\\
&&\quad+ \overline B(m^2,q^2)\left(n^2-4\right)\Bigg[
  \frac{8q^2}{n}L^r_{5}+M^2_M\Bigg(\frac{2}{n^2}L
- 16 L^r_{4}
-{32\over n} L^r_{5}  -32 L^r_{6}  +{64\over n} L^r_{8}\Bigg) \Bigg]
\nonumber\\
&&\quad+\overline B(m^2,q^2)^2
\left(n^2-4\right)\left(\frac{q^2}{4}-\frac{2M^2_M}{n^2}\right) \Bigg\}\,,
\\
%%%%%%%%%%%%%%%%%%%%%%%%%%%
&&\mathbf{Real}\nonumber\\
\Pi_{SS}&=&B^2_0\left\{  8 H^r_2+16 L^r_{8} 
+\frac{1}{n}\left(n-1\right)\left(n+2\right)
 \overline B(m^2,q^2)\right\}
\nonumber\\
&&+{B^2_0\over F^2_M }\Bigg\{
q^2 \left(8 K^r_{113}+32 K^r_{47}\right)
+ M^2_M \Bigg(192 K^r_{25}+128 n K^r_{26}\Bigg)
\nonumber\\&&
\quad
+M^2_M L\Bigg[\left(\frac{32}{n}-32-32n\right) L^r_{8}
-64 L^r_{7}
+\left(n-1\right)\left(n+2\right)\frac{16}{n} L^r_{5}\Bigg]
\nonumber\\
&&\quad+ \overline B(m^2,q^2)\left(n-1\right)\left(n+2\right)\Bigg[
  \frac{8q^2}{n}L^r_{5}+M^2_M\Bigg(\left(-\frac{1}{n}+\frac{1}{n^2}\right)L
- 32 L^r_{4}
\nonumber\\
&&\quad
-{32\over n} L^r_{5}  +64 L^r_{6}  +{64\over n} L^r_{8}\Bigg) \Bigg]
\nonumber\\
&&\quad
+\overline B(m^2,q^2)^2
\left(n-1\right)\left(n+2\right)\left[\frac{q^2}{4}
+M^2_M\left(\frac{1}{2n}-\frac{1}{n^2}\right)\right] \Bigg\}\,,
\\
%%%%%%%%%%%%%%%%%%%%%%%%%%%
&&\mathbf{Pseudo-Real}\nonumber\\
\Pi_{SS}&=&B^2_0\left\{  8 H^r_2+16 L^r_{8} 
+\frac{1}{n}\left(n+1\right)\left(n-2\right)
 \overline B(m^2,q^2)\right\}
\nonumber\\
&&
+{B^2_0\over F^2_M }\Bigg\{
q^2 \left(8 K^r_{113}+32 K^r_{47}\right)
+ M^2_M \Bigg(192 K^r_{25}+128 n K^r_{26}\Bigg)
\nonumber\\&&
\quad\quad
+M^2_M L\Bigg[\left(\frac{32}{n}+32-32n\right) L^r_{8}
-64 L^r_{7}
+\left(n+1\right)\left(n-2\right)\frac{16}{n} L^r_{5}\Bigg]
\nonumber\\
&&\quad+ \overline B(m^2,q^2)\left(n+1\right)\left(n-2\right)\Bigg[
  \frac{8q^2}{n}L^r_{5}+M^2_M\Bigg(\left(\frac{1}{n}+\frac{1}{n^2}\right)L
- 32 L^r_{4}
\nonumber\\
&&\quad\quad
-{32\over n} L^r_{5}  +64 L^r_{6}  +{64\over n} L^r_{8}\Bigg) \Bigg]
\nonumber\\
&&\quad+\overline B(m^2,q^2)^2
\left(n+1\right)\left(n-2\right)\left[\frac{q^2}{4}
+M^2_M\left(-\frac{1}{2n}-\frac{1}{n^2}\right)\right] \Bigg\}\,.
\ea
The definition of the one-loop function $\overline B(m^2,q^2)$
can be found in Appendix~\ref{oneloopint}.

\subsubsection{Singlet case}

We have also calculated the singlet case. This is the quark-antiquark
combination that shows up in the mass term.

We write the expression up to NNLO as:
%\begin{eqnarray}
%\Pi^0_{SS} = B^2_0 \Pi^0_{SS}(NLO) + \frac{B^2_0}{F^2_M}  \Pi^0_{SS}(NNLO)
%\end{eqnarray}
\ba
&&\mathbf{Complex}\nonumber\\
\Pi^0_{SS}&=&B^2_0 \Bigg\{8 n H^r_2 +32 n^2 L^r_6 +16 n L^r_8 
+2(n^2-1)\overline B(m^2,q^2) \Bigg\}
\nonumber\\
&&+\frac{B^2_0}{F^2_M}\Bigg\{
8q^2 \left( n K^r_{113}+ 4n  K^r_{47} +4n^2 K^r_{48} \right)
+192 M^2_M \left(nK^r_{25} + n^2K^r_{26}+ n^3K^r_{27} \right)
\nonumber\\&&
\quad
+ M^2_M L\left(n^2-1\right)
\left(32 n L^r_4  +32 L^r_5 - 64 n L^r_6 -64  L^r_8 \right)
\nonumber\\
&&\quad
 + \overline B(m^2,q^2) \left(n^2-1\right)
\Bigg[16 q^2 \left(nL^r_4 + L^r_5\right)
\nonumber\\&&\quad\quad
 + M^2_M\left(\frac{4}{n} L
 +64 \left(2L^r_8  + 2nL^r_6  - L^r_5 - nL^r_4\right)\right)
\Bigg]
\nonumber\\
&&\quad + \overline B(m^2,q^2)^2\left(n^2-1\right)
  \left( n q^2- {2M^2_M\over n} \right)\Bigg\}\,,
\\
%%%%%%%%%%%%%%%%%%%%%%
&&\mathbf{Real}\nonumber\\
\Pi^0_{SS}&=&B^2_0 \Bigg\{16 n H^r_2 +128 n^2 L^r_6 +32 n L^r_8 
+2(2n^2+n-1)\overline B(m^2,q^2) \Bigg\}
\nonumber\\
&&+\frac{B^2_0}{F^2_M}\Bigg\{
16q^2 \left( n K^r_{113}+ 4n  K^r_{47} +8n^2 K^r_{48} \right)
+384 M^2_M \left(nK^r_{25} +2 n^2K^r_{26}+4 n^3K^r_{27} \right)
\nonumber\\&&
\quad
+ M^2_M L\left(2n^2+n-1\right)
\left(64 n L^r_4 + 32 L^r_5 - 128 n L^r_6 -64  L^r_8 \right)
\nonumber\\
&&\quad
 + \overline B(m^2,q^2) \left(2n^2+n-1\right)
\Bigg[16 q^2 \left(2nL^r_4 + L^r_5\right)
\nonumber\\&&\quad\quad
 + M^2_M \left(\left(-2+\frac{2}{n}\right) L
 +64 (2L^r_8  + 4nL^r_6  - L^r_5 - 2nL^r_4)\right)
\Bigg]
\nonumber\\
&&\quad + \overline B(m^2,q^2)^2\left(2n^2+n-1\right)
  \left[ n q^2+ M^2_M\left(1-{1\over n} \right)\right]\Bigg\}\,,\\
%%%%%%%%%%%%%%%%%%%%%%
&&\mathbf{Pseudo-Real}\nonumber\\
\Pi^0_{SS}&=&B^2_0 \Bigg\{16 n H^r_2 +128 n^2 L^r_6 +32 n L^r_8 
+2(2n^2-n-1)\overline B(m^2,q^2) \Bigg\}
\nonumber\\
&&+\frac{B^2_0}{F^2_M}\Bigg\{
16q^2 \left( n K^r_{113}+ 4n  K^r_{47} +8n^2 K^r_{48} \right)
+384 M^2_M \left(nK^r_{25} +2 n^2K^r_{26}+4 n^3K^r_{27} \right)
\nonumber\\&&
\quad
+ M^2_M L\left(2n^2-n-1\right)
\left(64 n L^r_4 + 32 L^r_5 - 128 n L^r_6 -64  L^r_8 \right)
\nonumber\\
&&\quad
 + \overline B(m^2,q^2) \left(2n^2-n-1\right)
\Bigg[16 q^2 \left(2nL^r_4 + L^r_5\right)
\nonumber\\&&\quad\quad
 + M^2_M \left(\left(2+\frac{2}{n}\right) L
 +64 (2L^r_8  + 4nL^r_6  - L^r_5 - 2nL^r_4)\right)
\Bigg]
\nonumber\\
&&\quad + \overline B(m^2,q^2)^2\left(2n^2-n-1\right)
  \left[ n q^2+ M^2_M\left(-1-{1\over n} \right)\right]\Bigg\}\,.
\ea
We also written the result in term of physical $M^2_M$ and $F_M$.
Notice that all loop diagrams are proportional to the number of Goldstone bosons
in each case, i.e. $n^2-1$, $2n^2+n-1$, $2n^2-n-1$ for the
complex, real and pseudo-real case respectively.

\subsection{The Pseudo-Scalar Two-Point Functions}
\label{sect:PP}

The pseudo-scalar two-point function is defined in (\ref{deftwop}).
Just as in the case of the axial-vector two-point function there are
one-particle-reducible diagrams. The diagrams are the same as those
for the axial-vector two-point function with the axial-vector current
replaced by a pseudo-scalar current. 
These are shown in Figure~\ref{aa2pointp2p4} and \ref{aa2pointp6}.
There is also no vertex with two
pseudo-scalar currents at LO so the equivalent of
diagrams (1) and (7) in Figure \ref{aa2pointp2p4}
and (13--15) in Figure \ref{aa2pointp6} vanish immediately.
Just as in the scalar case, one should distinguish here between two cases:
The adjoint case for the complex representation case which generalizes to the
broken generators for the real and pseudo-real case, and the singlet
operator with $T^a$ in (\ref{pseudoscalarcurrent}) the unit operator.

In Section \ref{sect:AA} we could simplify the final expressions very much
by writing the final expression with the single pole at the meson mass
in terms of the decay constant. The same happens here if we instead
rewrite the result in terms of the meson pseudo-scalar decay constant $G_M$.
So we first need to obtain that quantity to NNLO.

\subsubsection{The meson pseudo-scalar decay constant $G_M$}

The decay constant of the
 pseudoscalar density to the mesons, $G_M$
is defined\footnote{The $\sqrt{2}$ is included in the definition to have the
same normalization as \cite{GL1}.} similarly to $F_M$:
\be
\langle 0 |\bar q i\gamma_5 T^a q | \pi^b \rangle = \frac{1}{\sqrt2}\delta^{ab} G_M
\ee
The calculation of $G_M$ is very similar to $F_M$, the diagrams are exactly
those shown in Figure 2 in \cite{paper1} with one of the legs replaced
by the pseudo-scalar current. There is here also a contribution
from wave-function renormalization.
In \cite{paper1} we reported all the quantities $M_M^2$, $F_M$
and $\langle\bar q q\rangle$ as an expansion in the bare or lowest
order quantities $F$ and $M^2=2B_0 m$. We therefore do the same here.
We therefore use the quantity
\be
L_0 = \frac{1}{16\pi^2} \log\frac{M^2}{F^2}
\ee
instead of $L$ as in the other sections of this paper.

This quantity has been calculated to NLO in two-flavour ChPT in \cite{GL1}
and was  called $G_\pi$ there.
We have checked that our NLO result agrees with theirs.

At leading order, all the three case have same expression:
\be
G_M = G_0 = 2 B_0 F\,.
\ee
We express the full results up to NNLO in terms of the LO meson mass $M^2$
and decay constant $F$ as
\be
G_M = 2 B_0 F \left(1+\frac{M^2}{F^2}\ a_G + \frac{M^4}{F^4}\ b_G
 \right)
\ee
At NLO and NNLO, the coefficients $a_G$ and $b_G$ are
\ba
&&\mathbf{Complex}\nonumber\\
 a_G &=& \left(\frac{1}{n}-\frac{n}{2}\right)L_0
+4(-n L^r_4-L^r_5+ 4 nL^r_6+4 L^r_8) \nonumber\\
 b_G &=&
-64(L^r_5 +nL^r_4)(L^r_8+nL^r_6)+24(L^r_5+nL^r_4)^2\nonumber\\
&&-8 n^2 K^r_{22}+48 n^2 K^r_{27}-32 K^r_{17}-8 K^r_{19}-8 K^r_{23}+48 K^r_{25}+32 K^r_{39}-32 n K^r_{18} \nonumber\\
&&-8 n K^r_{20}-8 n K^r_{21}+48 n K^r_{26}+32 n K^r_{40}   \nonumber\\
&&+L_0\Bigg[-(32-22n^2)\left(L^r_4+{1\over n}L^r_5\right) + (4-8n^2)(L^r_1+4L^r_6)\nonumber\\
&&\qquad+\left({80\over n} - 48 n\right)L^r_8+\left({12\over n} - 10 n\right)L^r_3
-\left(8 +2 n^2\right)L^r_2 +\left({12\over n} - 4 n\right)L^r_0 \Bigg]\nonumber\\
&&+\pi_{16}\Bigg[(2-n^2)\left({8\over n}L^r_8 + 8L^r_6 - {4\over n}L^r_5 - 4L^r_4
 -{1\over n}L^r_3\right)\nonumber\\
&&\qquad  +n^2L^r_2 +2L^r_1+2\left(n-{1\over n}\right) L^r_0\Bigg] \nonumber\\
&&+\pi^2_{16}\left(\frac{113 n^2}{256}-\frac{13}{24}+\frac{13}{8 n^2}\right)
- \pi_{16} L_0\left(\frac{55 n^2}{96}-1+\frac{7}{2 n^2}\right)\nonumber\\
&&+L_0^2 \left(\frac{3 n^2}{16}-\frac{3}{2}+\frac{9}{2 n^2}\right)\,,
\\
%%%%%%%%%%%%%%%%%%%%%%%%%%%%
&&\mathbf{Real}\nonumber\\
 a_G &=&
-\left(\frac{n}{2}+\frac{1}{2}-\frac{1}{2 n}\right)L_0
+ (-8n L^r_4 -4 L^r_5+32 nL^r_6+16 L^r_8) \nonumber\\
 b_G &=&
-64(L^r_5 + 2nL^r_4)(L^r_8+ 2nL^r_6)+24(L^r_5+ 2nL^r_4)^2\nonumber\\
&&-32 K^r_{22} n^2+192 K^r_{27} n^2 -32 K^r_{17}-8 K^r_{19}-8 K^r_{23}+48 K^r_{25}+32 K^r_{39}\nonumber\\
&&-64 K^r_{18} n-16 K^r_{20} n-16 K^r_{21} n+96 K^r_{26} n+64 K^r_{40} n\nonumber\\
&&+L_0\Bigg[(-16+16n+22n^2)\left(2L^r_4+{1\over n}L^r_5\right) + (4-8n-16n^2)L^r_1\nonumber\\
&&\qquad
+ 16(1-3n-4n^2)L^r_6
-\left(40-{40\over n} + 48 n\right)L^r_8-\left(6-{6\over n} + 10 n\right)L^r_3\nonumber\\
&&\qquad-\left(8 +2n +4n^2\right)L^r_2 -\left(6-{6\over n} + 4 n\right)L^r_0 \Bigg]\nonumber\\
&&+\pi_{16}\Bigg[(1-n-n^2)\left({8\over n}L^r_8 + 16L^r_6 - {4\over n}L^r_5 - 8L^r_4 -{1\over n}L^r_3\right)\nonumber\\
&& \qquad +(n+2n^2)L^r_2 +2L^r_1+ \left(1-{1\over n}+2n\right) L^r_0\Bigg]\nonumber\\
&&+\pi^2_{16}\left(\frac{113 n^2}{256}+\frac{443 n}{768}-\frac{13}{96}-\frac{13}{32 n}+\frac{13}{32 n^2}\right) \nonumber\\
&&-\pi_{16} L_0 \left(\frac{55 n^2}{96}+\frac{67 n}{96}-\frac{3}{8}-\frac{5}{8 n}+\frac{7}{8 n^2}\right)\nonumber\\
&&+ L_0^2\left(\frac{3 n^2}{16}+\frac{13 n}{16}-\frac{1}{8}-\frac{3}{2 n}+\frac{9}{8 n^2}\right)\,,
\\
%%%%%%%%%%%%%%%%%%%%%%%%%%%%
&&\mathbf{Pseudo-Real}\nonumber\\
 a_G&=& - L_0 \left(\frac{n}{2}-\frac{1}{2}-\frac{1}{2 n}\right)
+ (-8n L^r_4 -4 L^r_5+32 nL^r_6 +16 L^r_8) \nonumber\\
 b_G&=&-64(L^r_5 + 2nL^r_4)(L^r_8+ 2nL^r_6)+24(L^r_5+ 2nL^r_4)^2\nonumber\\
&&-32 K^r_{22} n^2+192 K^r_{27} n^2 -32 K^r_{17}-8 K^r_{19}-8 K^r_{23}+48 K^r_{25}\nonumber\\
&&+32 K^r_{39}-64 K^r_{18} n-16 K^r_{20} n-16 K^r_{21} n+96 K^r_{26} n+64 K^r_{40} n\nonumber\\
&&+L_0\Bigg[(-16-16n+22n^2)\left(2L^r_4+{1\over n}L^r_5\right) 
+ (4+8n-16n^2)L^r_1
\nonumber\\
&&\qquad
 + 16(1+3n-4n^2)L^r_6
+\left(40+{40\over n} - 48 n\right)L^r_8+\left(6+{6\over n} - 10 n\right)L^r_3\nonumber\\
&&\qquad-\left(8 -2n +4n^2\right)L^r_2 +\left(6+{6\over n} - 4 n\right)L^r_0 \Bigg]\nonumber\\
&&+\pi_{16}\Bigg[(1+n-n^2)\left({8\over n}L^r_8 + 16L^r_6 - {4\over n}L^r_5 - 8L^r_4 -{1\over n}L^r_3\right)\nonumber\\
&& \qquad +(-n+2n^2)L^r_2 +2L^r_1+ \left(-1-{1\over n}+2n\right) L^r_0\Bigg]\nonumber\\
&&+ \pi^2_{16}\left(\frac{113 n^2}{256}-\frac{443 n}{768}-\frac{13}{96}+\frac{13}{32 n}
+\frac{13}{32 n^2}\right) \nonumber\\
&&- \pi_{16}L_0\left(\frac{55 n^2}{96}-\frac{67
n}{96}-\frac{3}{8}+\frac{5}{8 n}
+\frac{7}{8 n^2}\right)\nonumber\\
&&+L_0^2 \left(\frac{3 n^2}{16}-\frac{13 n}{16}-\frac{1}{8}+\frac{3}{2
n}+\frac{9}{8 n^2}\right)\,.
\ea

\subsubsection{$X^a$ case}

The pseudo-scale two point functions are similar to the axial-vector ones
in the diagrams as described above.
The LO result is the same
for all the three cases:
\be
\Pi^a_{PP} = -{1\over2}{ G^2_0\over q^2-M^2}\,.
\ee
The superscript ``$a$'' indicates the case with $T^a$ in
(\ref{pseudoscalarcurrent}) an $SU(n)$ generator. For the real and
pseudo-real case this is related by the conserved part of the symmetry group
also to a number of diquark currents.

Subtracting the pole contribution in terms of the physical
mass and decay constants, $M^2_M$, $F_M$ and $G_M$,
absorbs the major part of the higher order corrections.
The final results are thus much simpler when written in this way.
The remaining part at NLO is
\be
\Pi^a_{PP} = B^2_0 (8 H^r_2-16 L^r_8)\,.
\ee
 Thus we can define the full NNLO results as
\be
\Pi^a_{PP} = -{1\over2}{ G_M^2\over q^2-M^2_M} + B^2_0 (8 H^r_2-16 L^r_8)
+ \frac{ B^2_0}{F^2_M}\,\hat \Pi^a_{PP}\,,
\ee
where the $ \hat \Pi_{PP}$ is the remainder at NNLO.
Its expression for the three different cases is:
\ba
&&\mathbf{Complex} \nonumber\\
\hat\Pi^a_{PP}&=&
- {3\over2}n^2 q^4 H^M_{21}(M^2_M,M^2_M,M^2_M,q^2)\nonumber\\
&&
+\left[\left(\frac{4}{3}-\frac{4}{n^2}\right) q^4
-{n^2\over2} M^2_M  q^2\right] H^M(M^2_M,M^2_M,M^2_M,q^2)\nonumber\\
&&
+ 8q^2K^r_{113}
+ 64M^2_M (K^r_{17}+ nK^r_{18}  -K^r_{39}- nK^r_{40} )\nonumber\\
&&
+ L^2 M^2_M \left(-\frac{n^2}{2}-\frac{6}{n^2}+2\right)
+ L M^2_M \left(64 L^r_6+32 nL^r_8  -\frac{64 L^r_8}{n}\right)\nonumber\\
&&
+ \pi_{16}L\left[ M^2_M \left(-\frac{8}{3}-\frac{n^2}{12}+\frac{8}{n^2}\right)
+ \left(\frac{2}{3}-\frac{n^2}{8}-\frac{2}{n^2}\right)  q^2\right]\nonumber\\
&&
+ \pi_{16}^2\left[M^2_M \left(\frac{5}{3}-\frac{85 n^2}{96}-\frac{5}{n^2}\right)
+ q^2\left(\frac{7}{6}-\frac{15 n^2}{64}-\frac{7}{2 n^2}\right)  \right] \,,
\\
%%%%%%%%%%%%%%%%%%%%%%%
&&\mathbf{Real} \nonumber\\
\hat\Pi^a_{PP}&=&
  -\frac{3}{2}q^4 n(n+ 1) H^M_{21}(M^2_M,M^2_M,M^2_M,q^2) \nonumber\\
&&+ \left[ \left(-\frac{1}{n^2}-\frac{n}{3}+\frac{1}{n}+\frac{1}{3}\right) q^4
-  \frac{1}{2}M^2_M q^2 n\left(n+1\right) \right] H^M(M^2_M,M^2_M,M^2_M,q^2) \nonumber\\
&&+8 q^2   K^r_{113}
+  64 M^2_M (K^r_{17}+2n K^r_{18}-K^r_{39}-2n K^r_{40}) \nonumber\\
&&+   M^2_M L^2\left(-\frac{n^2}{2}-\frac{3}{2 n^2}-n+\frac{3}{2 n}+\frac{1}{2}\right)
+32 M^2_M L \left(2L^r_6+L^r_8 n-\frac{L^r_8}{n}+L^r_8\right) \nonumber\\
&&+  \pi_{16}L \left[ M^2_M \left(-\frac{n^2}{12}+\frac{2}{n^2}+\frac{7 n}{12}-\frac{2}{n}-\frac{2}{3}\right)
+ q^2\left(-\frac{n^2}{8}-\frac{1}{2 n^2}-\frac{7 n}{24}+\frac{1}{2 n}+\frac{1}{6}\right)\right]\nonumber\\
&&+  \pi_{16}^2\Bigg[ M^2_M \left(-\frac{85 n^2}{96}-\frac{5}{4 n^2}-\frac{125 n}{96}+\frac{5}{4 n}+\frac{5}{12}\right)\nonumber\\
&&\quad  +q^2\left(-\frac{15 n^2}{64}-\frac{7}{8 n^2}-\frac{101 n}{192}+\frac{7}{8 n}+\frac{7}{24}\right)\Bigg]\,,
\\
%%%%%%%%%%%%%%%%%%%%%%%
&&\mathbf{Pseudo-Real} \nonumber\\
\hat\Pi^a_{PP}&=&
 \frac{3}{2}  q^4 n(1-n)H^M_{21}(M^2_M,M^2_M,M^2_M,q^2) \nonumber\\
&&+ \left[\left(-\frac{1}{n^2}+\frac{n}{3}-\frac{1}{n}+\frac{1}{3}\right) q^4
+ \frac{1}{2}M^2_M  q^2n(1-n)\right] H^M(M^2_M,M^2_M,M^2_M,q^2) \nonumber\\
&&+8q^2 K^r_{113}
+ 64M^2_M (K^r_{17}+2nK^r_{18}-K^r_{39}-2nK^r_{40}) \nonumber\\
&&+M^2_M  L^2  \left(-\frac{n^2}{2}-\frac{3}{2 n^2}+n-\frac{3}{2 n}+\frac{1}{2}\right)
+ M^2_M L\left[64 L^r_6+32  \left(n-\frac{1}{n}-1\right) L^r_8\right]\nonumber\\
&&+\pi_{16}L\Bigg[  M^2_M \left(-\frac{n^2}{12}+\frac{2}{n^2}-\frac{7 n}{12}+\frac{2}{n}-\frac{2}{3}\right)
+ q^2 \left(-\frac{n^2}{8}-\frac{1}{2 n^2}+\frac{7 n}{24}-\frac{1}{2 n}+\frac{1}{6}\right)  \Bigg]\nonumber\\
&&+\pi_{16}^2 \Bigg[ M^2_M \left(-\frac{85 n^2}{96}-\frac{5}{4 n^2}+\frac{125 n}{96}-\frac{5}{4 n}+\frac{5}{12}\right)\nonumber\\
&&\quad+ \left(-\frac{15 n^2}{64}-\frac{7}{8 n^2}+\frac{101 n}{192}-\frac{7}{8 n}+\frac{7}{24}\right) q^2\Bigg]\,.
\ea
The loop integrals $H^M$ and $H^M_{21}$ are defined in Appendix \ref{sunsetint}.

\subsubsection{Singlet case}

In the singlet case with $a=0$, there is no contribution with poles.
Only the one-particle-irreducible diagrams contribute.
As a consequence, there is no order $p^2$ contribution and at order $p^4$
there is only a tree level contribution from the equivalent
of diagram (3) in Figure~\ref{aa2pointp2p4}.
At order $p^6$ or NNLO only the one-particle-irreducible diagrams
contribute and since there is no order $p^2$ vertex with two pseudo-scalar
currents only diagram (11--12) and (16) in Figure~\ref{aa2pointp6}
contribute.

Since there is no single pole contribution, there is also no need here to
expand in the integrals around the meson mass.
The integral $H^F$ is defined in Appendix \ref{sunsetint}.

The singlet pseudo-scalar two-point function we write as
\be
\Pi^0_{PP} = B^2_0 \overline\Pi^0_{PP}+\frac{B_0^2}{F^2_M}\hat\Pi^0_{PP}\,.
\ee
The results for the three cases are
\ba
&&\mathbf{Complex}: \nonumber\\
\overline\Pi^0_{PP}&=& 8 n H^r_2-16 n L^r_8-32 n^2 L^r_7\,,
\nonumber\\
 \hat\Pi^0_{PP}&=&
-\frac{2}{3n}\left(n^2-1\right)\left(n^2-4\right) H^F(M^2_M,M^2_M,M^2_M,q^2) \nonumber\\
&&+  q^2 \left(8 K^r_{113} n-32 K^r_{46} n^2\right)
-64M^2_M \left( K^r_{39} n+ K^r_{40} n^2+ K^r_{41} n^2+ K^r_{42} n^3\right) \nonumber\\&&
+  L^2 M^2_M \frac{1}{n}\left(n^2-1\right)\left(n^2-4\right)
+  64L M^2_M  (n^2-1)(  nL^r_7+  L^r_8) \nonumber\\&&
+ M^2_M \pi_{16}^2 \frac{1}{n}\left(n^2-1\right)\left(n^2-4\right)
\left(\frac{\pi^2}{6}+1\right)\,,
\\
%%%%%%%%%%%%%%%%%%%%%%
&&\mathbf{Real}:\nonumber\\
\overline\Pi^0_{PP}&=& 16n H^r_2 -128n^2 L^r_7 -32n L^r_8\,,
\nonumber\\
 \hat\Pi^0_{PP}&=&
 -\frac{2}{3n}\left(2n^2+n-1\right)\left(n^2+n-2\right)
 H^F(M^2_M,M^2_M,M^2_M,q^2) \nonumber\\
&&+  q^2 \left(16 K^r_{113} n-128 K^r_{46} n^2\right)
- 128 M^2_M \left(nK^r_{39} +2n^2K^r_{40} +2 n^2K^r_{41}+4n^3 K^r_{42} \right) \nonumber\\
&&+  L^2 M^2_M\frac{1}{n}\left(2n^2+n-1\right)\left(n^2+n-2\right)
+  64 M^2_M L(2 n^2+ n-1)(2nL^r_7 + L^r_8) \nonumber\\
&&+  M^2_M \pi_{16}^2 \frac{1}{n}\left(2n^2+n-1\right)\left(n^2+n-2\right)
\left(\frac{\pi^2}{6}+1\right)\,,
\\
%%%%%%%%%%%%%%%%%%%%%%
&&\mathbf{Pseudo-Real}:\nonumber\\
\overline\Pi^0_{PP}&=& 16n H^r_2 -128n^2 L^r_7 -32n L^r_8\,,
\nonumber\\
 \hat\Pi^0_{PP}&=&
 -\frac{2}{3n}\left(2n^2-n-1\right)\left(n^2-n-2\right)
 H^F(M^2_M,M^2_M,M^2_M,q^2) \nonumber\\
&&+  q^2 \left(16 K^r_{113} n-128 K^r_{46} n^2\right)
- 128 M^2_M \left(nK^r_{39} +2n^2K^r_{40} +2 n^2K^r_{41}+4n^3 K^r_{42} \right) \nonumber\\
&&+  L^2 M^2_M\frac{1}{n}\left(2n^2-n-1\right)\left(n^2-n-2\right)
+  64 M^2_M L(2 n^2-n-1)(2nL^r_7 + L^r_8) \nonumber\\
&&+  M^2_M \pi_{16}^2 \frac{1}{n}\left(2n^2-n-1\right)\left(n^2-n-2\right)
\left(\frac{\pi^2}{6}+1\right)\,.
\ea
Notice that just as for the scalar singlet two-point function, 
all loop contributions are
proportional to the number of Goldstone bosons.

\subsection{Large $n$}

As one can see from all the explicit formulas, 
many of the expressions become equal for the different cases
in the large $n$ limit .

\section{The Oblique Corrections and S-parameter}
\label{sect:S}

The physical process at the CERN LEP collider is $e^+ + e^- \to q + \bar
q$. There are three types of one loop correction to this process:
vacuum polarization corrections, vertex corrections, and box
corrections. The vacuum polarization contribution is
independent of the external fermions and it dominates the contributions
from physics beyond SM. For the light fermions, the other two corrections are
suppressed by factor of $m^2_f/m^2_Z$. That's why the vacuum
polarization corrections are called ``oblique corrections,'', and the
vertex and box corrections are called ``nonoblique corrections.''

\begin{figure}
\centering
\includegraphics[width=0.4\textwidth]{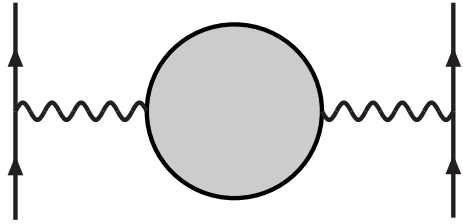}
\caption{The one-loop oblique correction to LEP process $e^+ + e^-
\to q + \bar q$.}
\label{oblique}
\end{figure}

The oblique polarization  only affect the gauge bosons propagators
and their mixing. The vacuum polarization amplitude can be defined
as 
\be
g^{\mu\nu}\Pi_{XY} + (q^\mu q^\nu\ \mathrm{terms}) = i\int
d^4x\; e^{iq\cdot x}\;\langle 0|T(J^\mu_X(x)J^\nu_Y(0))|0\rangle\,.
\ee
The influence of new physics to the oblique corrections can be
summarized to three parameters: $S$, $T$ and $U$. One can find their
definition in Ref. \cite{Peskin:1991sw}. These parameters are chosen
to be zero at a reference point in the SM. In the past 20 years,
they have been studied intensively in many models beyond the Standard
Model physics. 

For a beyond the Standard Model with strong dynamics at the TeV scale,
there will in general be many resonances and other nonperturbative effects.
At low momenta one can use the EFT as described above for these cases.
In this paper, we will estimate the $S$
parameter contribution from pseudo-Goldstone Boson sector
within the EFT. The
parameter $T$ and $U$  vanish because of the exact flavor
symmetry, i.e. we work in the equal mass case.

The $S$ parameter can be written as\footnote{Our two point functions are
normalized differently from those in \cite{Peskin:1991sw}.}
\cite{Peskin:1991sw}
\be
S=-2\pi \Big[ \Pi'_{VV}(0) -\Pi'_{AA}(0) \Big]
= 2\pi\frac{d}{dq^2}\left(q^2\Pi^{(1)}_{VV}-q^2\Pi^{(1)}_{AA}\right)_{q^2=0}\,.
\ee
$\Pi'_{VV}(0)$ and $\Pi'_{AA}(0)$ are the derivatives of the 
vector and axial-vector two-point functions at $q^2=0$.
One should keep in mind that $S$ is defined to be zero at a particular
place in the standard model, as discussed at the end of section V in 
\cite{Peskin:1991sw}. Our formulas are the equivalent of (5.12) in that
reference.

The result can be written as
\be
S = \overline S +\frac{\pi M^2_M}{F_M^2}\hat S\,,
\ee
with
\ba
&&\mathbf{Complex:}\nonumber\\
\overline S &=& -16\pi L^r_{10}-\frac{2n\pi}{3}\left(L+\pi_{16}\right)\,,
\nonumber\\
\hat S &=& 64\left( K^r_{102} - K^r_{81} + K^r_{17} + n K^r_{103} -  nf K^r_{82}
   +  n K^r_{18}\right)
  + \frac{n^2}{3} L^2 
\nonumber\\&&
  + 16 n \left(L^r_9+ 2 L^r_{10}\right)L
      - \pi_{16} \frac{11n^2}{9} L 
     +\pi_{16}^2 n^2 \left(\frac{85}{108} - \frac{5}{27}\tilde\psi\right)
\label{Scomplex}
\\
&&\mathbf{Real:}\nonumber\\
\overline S &=& -16\pi L^r_{10}-\frac{2(n+1)\pi}{3}\left(L+\pi_{16}\right)\,,
\nonumber\\
\hat S &=& 64\left( K^r_{102}- K^r_{81} + K^r_{17} +2 n K^r_{103} - 2 nf K^r_{82}
   + 2  n K^r_{18}\right)
  + \frac{n (n+1)}{3} L^2 
\nonumber\\&&
  + 16  \left[ (n+1) L^r_9+ (2 n+1) L^r_{10}\right]L
      - \pi_{16} \frac{11n(n+1)}{9} L 
\nonumber\\&&
     +\pi_{16}^2 n(n+1) \left(\frac{85}{108} - \frac{5}{27}\tilde\psi\right)
\,,
\label{Sreal}
\\
&&\mathbf{Pseudo-real:}\nonumber\\
\overline S &=& -16\pi L^r_{10}-\frac{2(n-1)\pi}{3}\left(L+\pi_{16}\right)\,,
\nonumber\\
\hat S &=& 64\left( K^r_{102}- K^r_{81} + K^r_{17} +2 n K^r_{103} - 2 nf K^r_{82}
   + 2  n K^r_{18}\right)
  + \frac{n (n-1)}{3} L^2 
\nonumber\\&&
  + 16  \left[ (n-1) L^r_9+ (2 n-1) L^r_{10}\right]L
      - \pi_{16} \frac{11n(n-1)}{9} L 
\nonumber\\&&
     +\pi_{16}^2 n(n-1) \left(\frac{85}{108} - \frac{5}{27}\tilde\psi\right)
\,.
\label{Spseudoreal}
\ea
The quantity $\tilde\psi$ is
\be
\tilde \psi = 6\sqrt{3} \mathrm{Cl}_2\left(\frac{2\pi}{3}\right)= 
7.0317217160684\,.
\ee

The real purpose of (\ref{Scomplex})-(\ref{Spseudoreal}) is to be able to
study the $S$-parameter in more general theories than just scaling
up from QCD. However to provide some feeling about numerical results
we choose parameters as if they are scaled up from QCD/ChPT.
We change $F_\pi=0.0922$~MeV to $F_M=243$~GeV and the subtraction scale from
$0.77~$GeV to $2$~TeV. We set the $K_i^r=0$ and keep $L_9^r=0.00593$
and $L_{10}^r= -0.00406$ 
at their values from ChPT \cite{BT,GPP}.

\begin{figure}
\centerline{
\begin{minipage}{0.41\textwidth}
\includegraphics[width=0.99\textwidth]{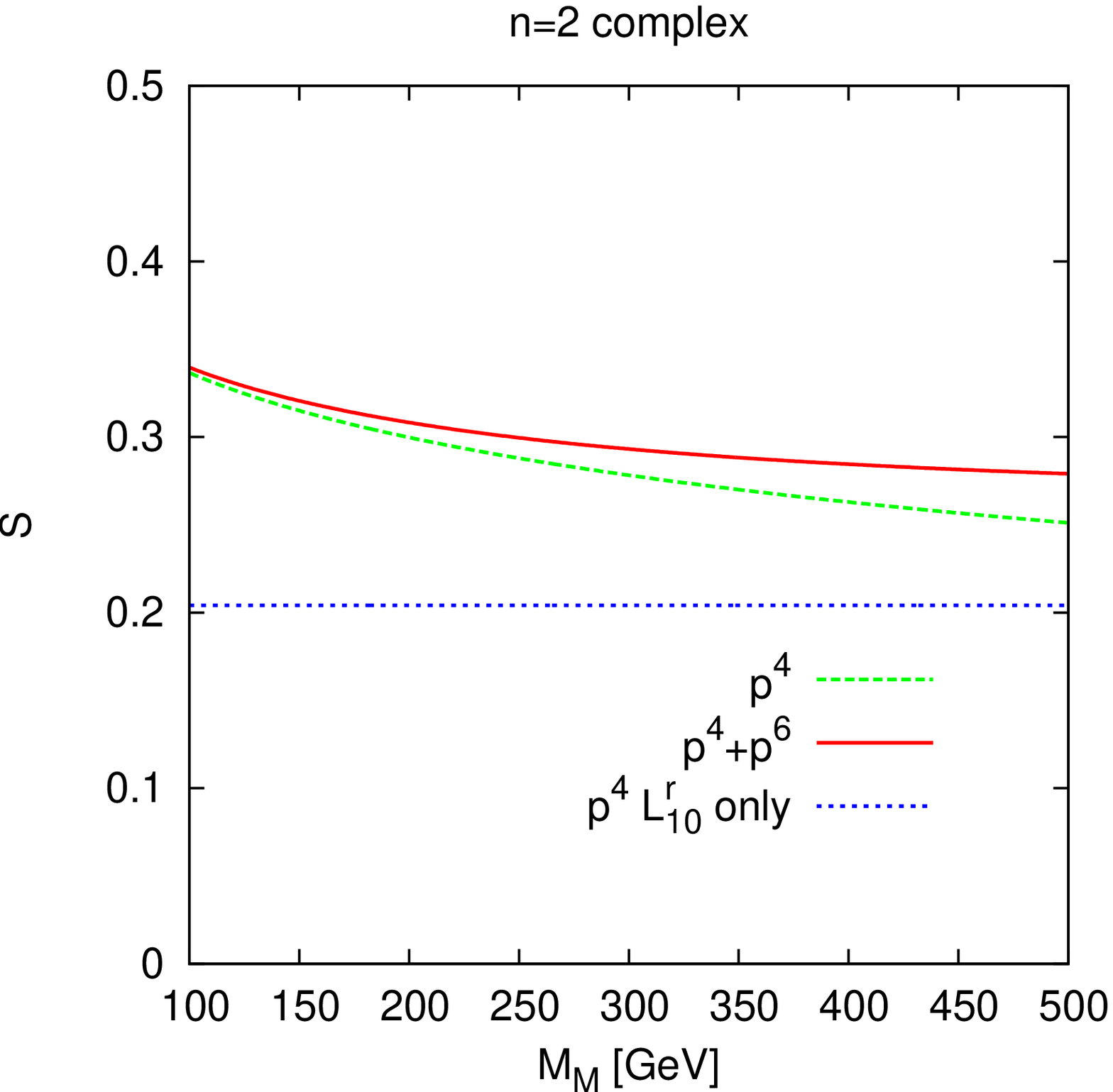}
\centerline{(a)}
\end{minipage}
~
\begin{minipage}{0.41\textwidth}
\includegraphics[width=0.99\textwidth]{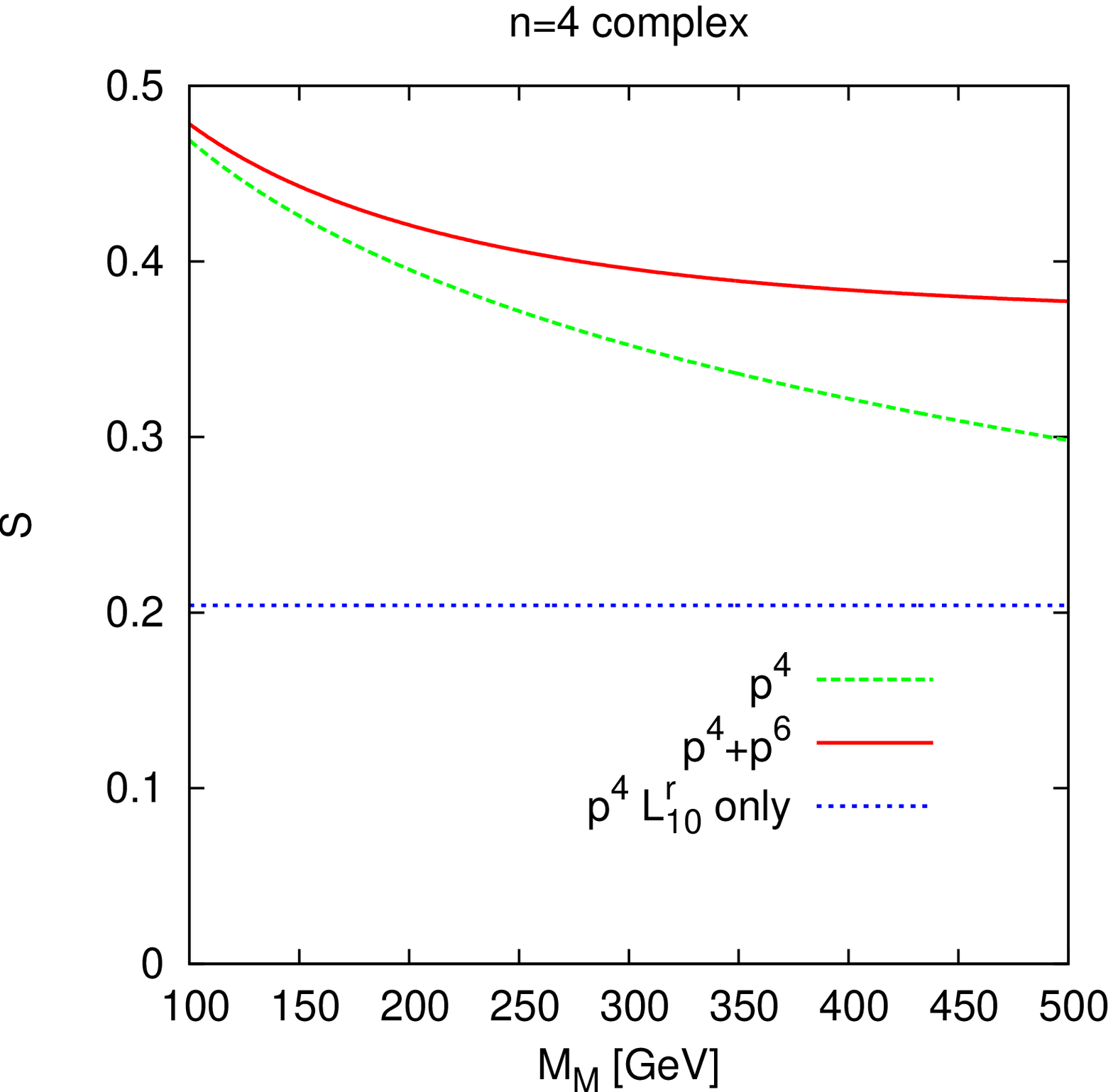}
\centerline{(b)}
\end{minipage}
}%centerline
\caption{\label{figScomplex} The $S$-parameter for
the values of $L_9^r$ and $L_{10}^r$ given in the text for the complex case.
(a) $n=2$ (b) $n=4$.}
\end{figure}

\begin{figure}
\centerline{
\begin{minipage}{0.41\textwidth}
\includegraphics[width=0.99\textwidth]{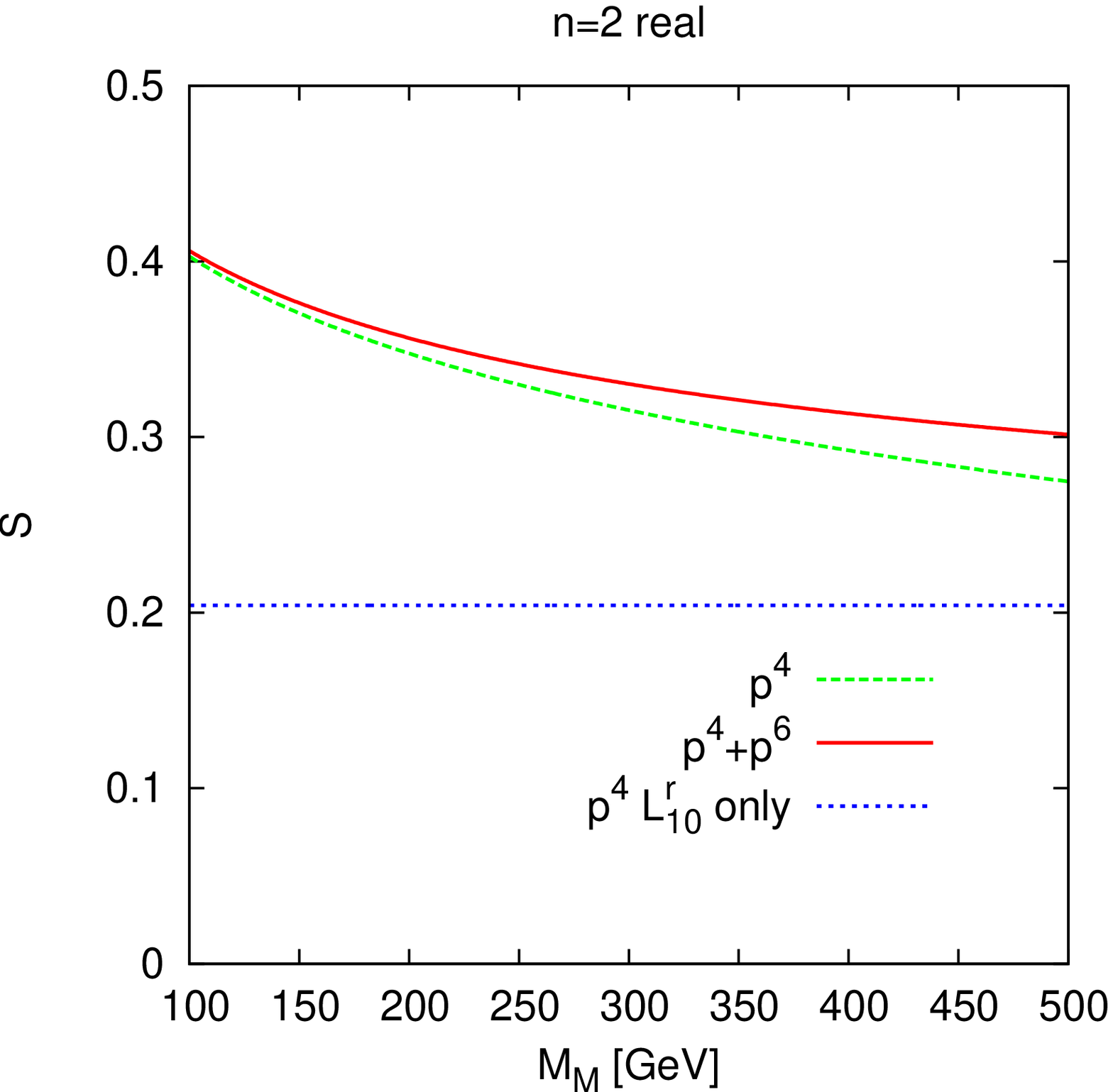}
\centerline{(a)}
\end{minipage}
~
\begin{minipage}{0.41\textwidth}
\includegraphics[width=0.99\textwidth]{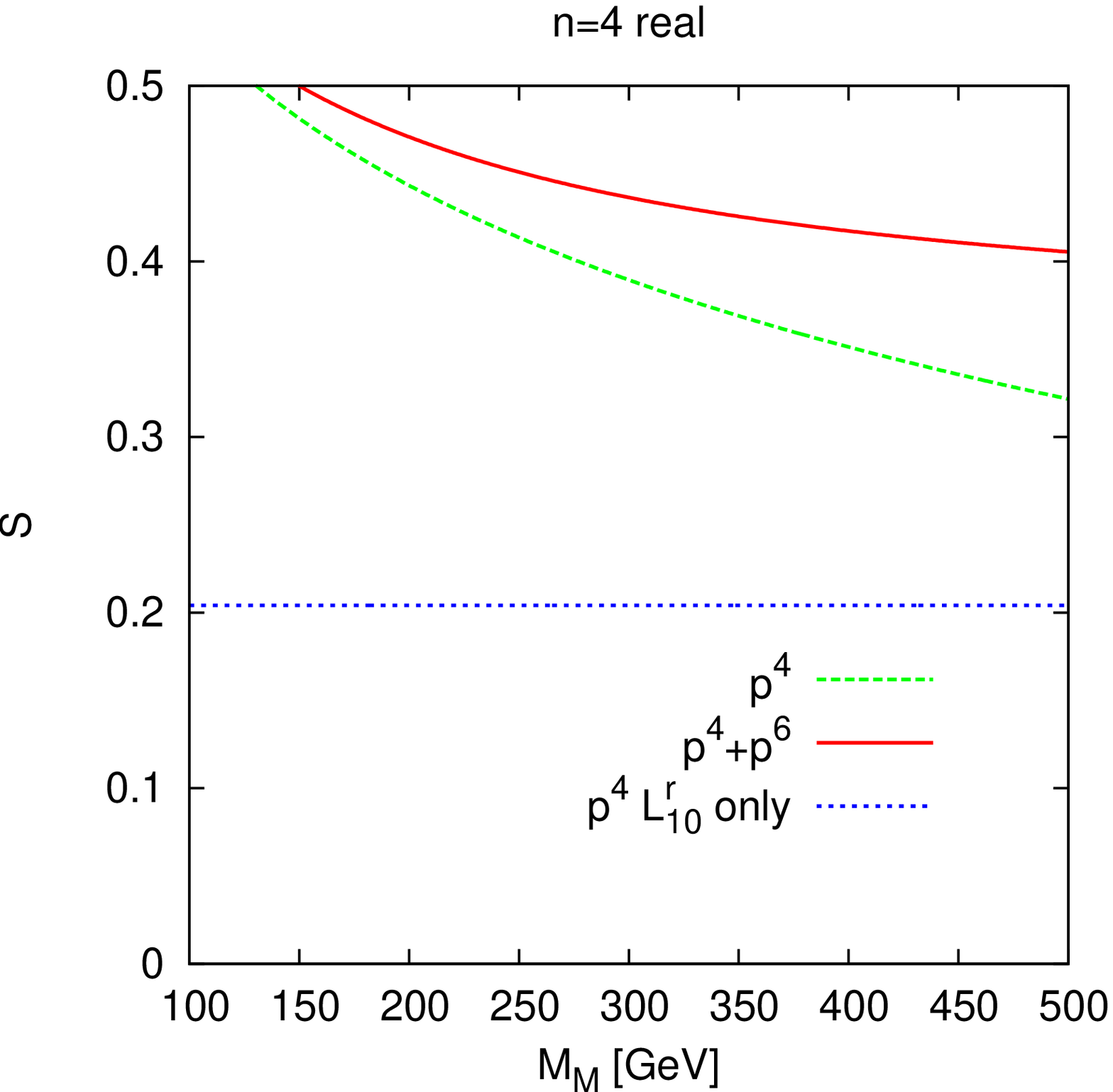}
\centerline{(b)}
\end{minipage}
}%centerline
\caption{\label{figSreal} The $S$-parameter for
the values of $L_9^r$ and $L_{10}^r$ given in the text for the real case.
(a) $n=2$ (b) $n=4$.}
\end{figure}

\begin{figure}
\centerline{
\begin{minipage}{0.45\textwidth}
\includegraphics[width=0.99\textwidth]{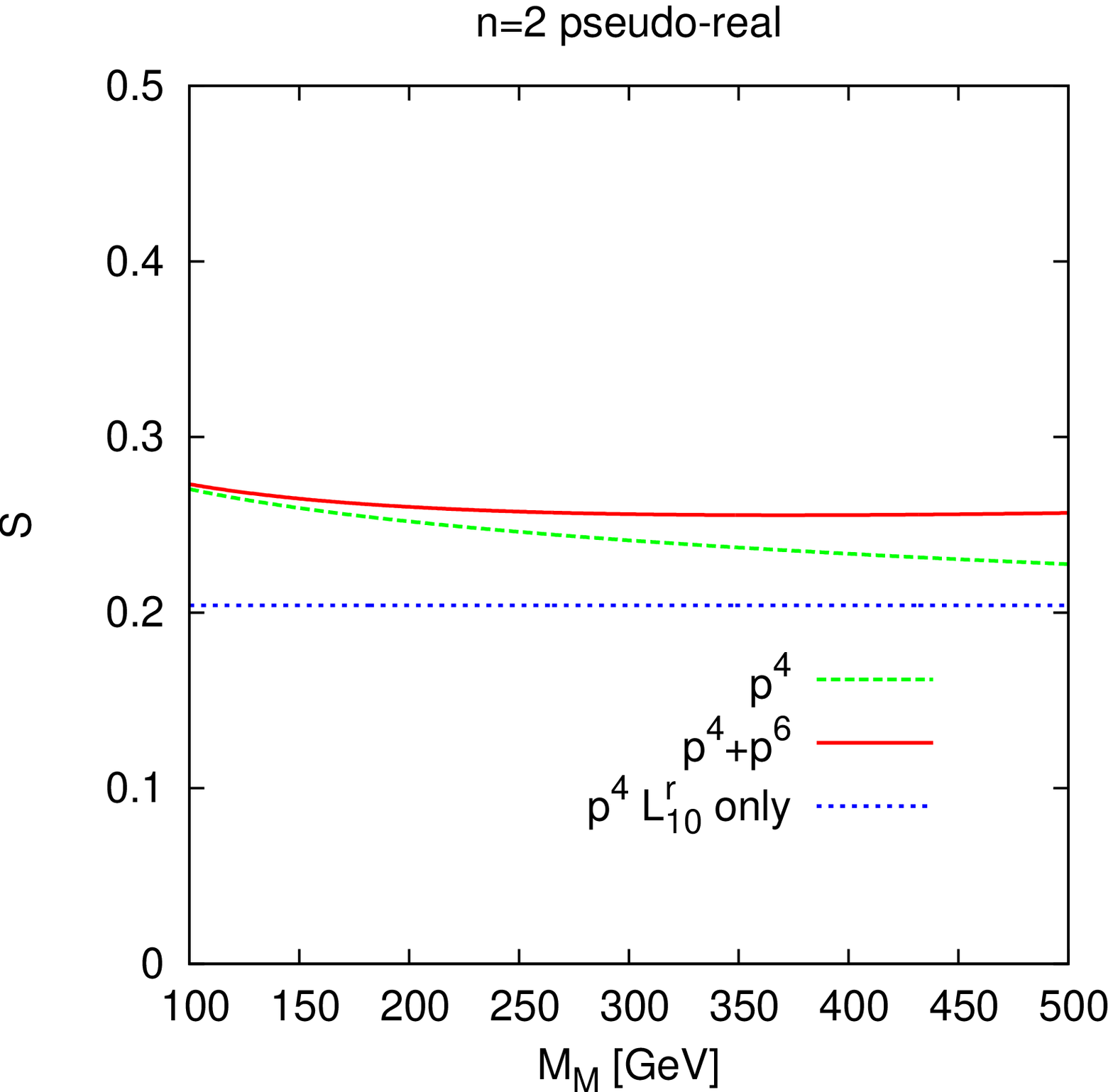}
\centerline{(a)}
\end{minipage}
~
\begin{minipage}{0.45\textwidth}
\includegraphics[width=0.99\textwidth]{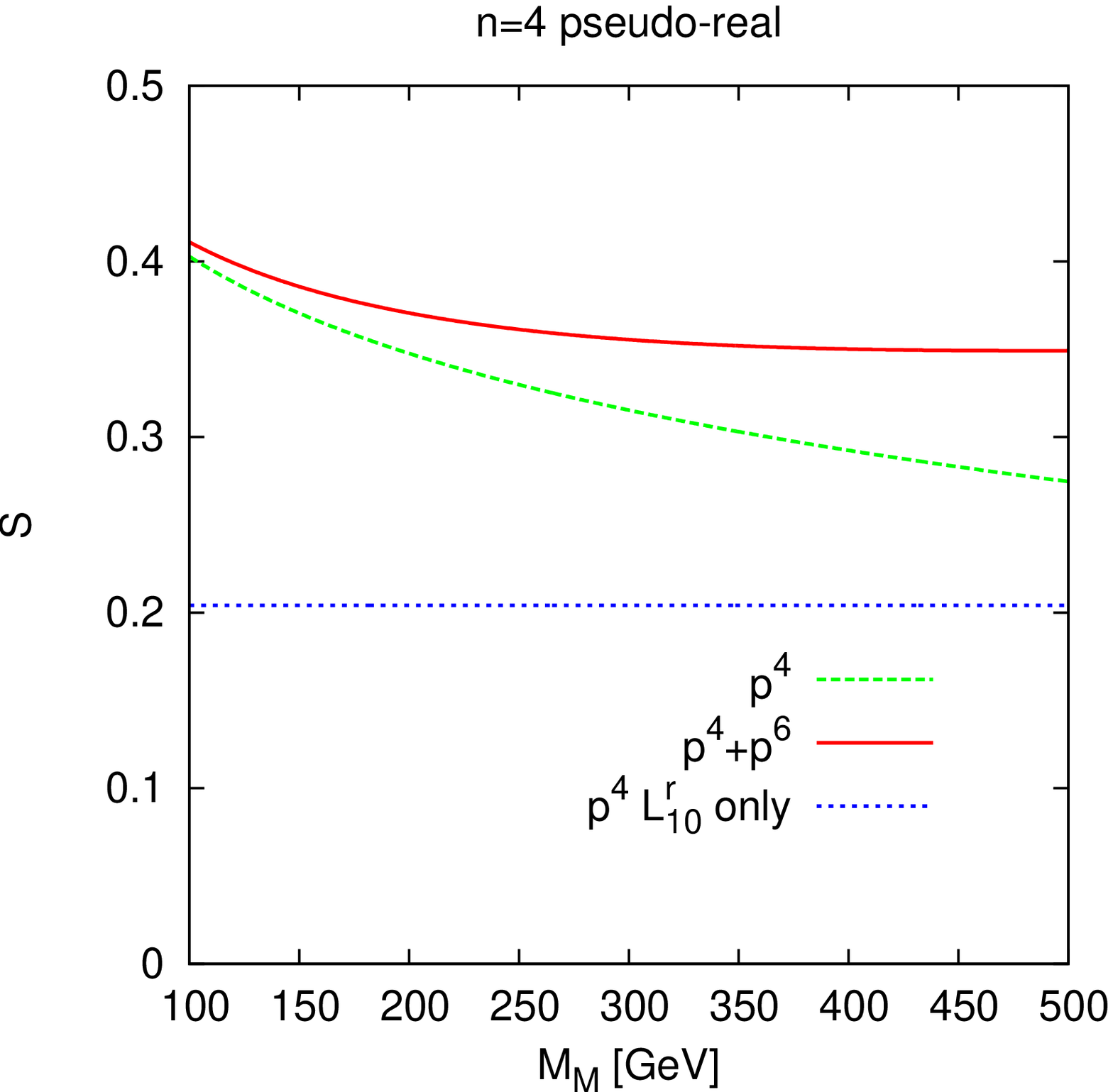}
\centerline{(b)}
\end{minipage}
}%centerline
\caption{\label{figSpseudoreal} The $S$-parameter for
the values of $L_9^r$ and $L_{10}^r$ given in the text for the pseudo-real case.
(a) $n=2$ (b) $n=4$.}
\end{figure}

In Figures~\ref{figScomplex}, \ref{figSreal} and \ref{figSpseudoreal}
we have shown the results for our three cases complex, real and pseudo-real
for $n=2$ and $n=4$. Shown are the full $p^4$ and $p^6$ contributions
as well as the $p^4$ part proportional to $L_{10}^r$ only.
The latter is what is the usual contribution to $S$ corrected for
the pieces that go into the reference point at $p^4$. We cannot do the same
for the full result since that depends on how one treats the extra
pseudo-Goldstone bosons that occur in the other models.


\section{Conclusion}
\label{sect:conclusions}

In this paper, we have calculated the two-point correlation
functions of vector,
axial-vector, scalar and pseudo-scalar currents for QCD-like
theories.

In the beginning of the paper, we gave a very brief overview of the QCD-like
theories and their EFT treatment as developed earlier.

We then gave the analytic results of those two-point functions up to NNLO.
The results are significantly shortened by using the physical meson mass
$M^2_M$ and decay constants $F_M$ and $G_M$ when rewriting the pole
contributions.

The main use of these formulas is expected to be in extrapolations to zero
fermion mass of technicolour related lattice calculations.
We have therefore also included precisely the combination needed for the
$S$-parameter.

\section*{Acknowledgments}

This work is supported in part by the European Community-Research
Infrastructure Integrating Activity ``Study of Strongly Interacting Matter'' 
(HadronPhysics2, Grant Agreement n. 227431)
and the Swedish Research Council grants 621-2008-4074 and 621-2010-3326.
This work used FORM \cite{FORM}.

\appendix
\section{Loop integrals}
\label{loopintegrals}

We use dimensional regularization and $\overline {MS}$ scheme to evaluate the loop integrals,  $d=4-2\epsilon$.

\subsection{One-loop integrals}
\label{oneloopint}

The  loop integral  with one propagator is
\ba
A(m^2)
&=& \frac{1}{i}\int \frac{d^d q}{(2\pi)^d}\frac{1}{q^2-m^2}
\nonumber\\
&=&\!\frac{m^2}{16\pi^2}\left\{
\lambda_0-\ln(m^2)
+\epsilon\left[\frac{C^2}{2}+\frac{1}{2}+\frac{\pi^2}{12}
+\frac{1}{2}\ln^2(m^2)
 - C\ln(m^2)\right]\right\}\nonumber\\
 && +{\cal O}(\epsilon^2)\,.
\ea
Here
$$
C=\ln(4\pi)+1-\gamma \qquad \qquad
\lambda_0 = {1\over\epsilon}+C
$$
The extra $+1$ in $C$ is the ChPT version of $\overline{MS}$.

The loop integrals with two propagators are
\ba
B(m_1^2,m_2^2,p^2)&=& \frac{1}{i}\int\frac{d^dq}{(2\pi)^d}
\frac{1}{(q^2-m_1^2)((q-p)^2-m_2^2)}\,,
\nonumber\\
B^\mu(m_1^2,m_2^2,p)
&=& \frac{1}{i}\int\frac{d^dq}{(2\pi)^d}
\frac{q^\mu}{(q^2-m_1^2)((q-p)^2-m_2^2)}
\\
&=& p^\mu B_1(m_1^2,m_2^2,p^2)\,,
\nonumber\\
B^{\mu\nu}(m_1^2,m_2^2,p)
&=& \frac{1}{i}\int\frac{d^dq}{(2\pi)^d}
\frac{q^\mu q^\nu}{(q^2-m_1^2)((q-p)^2-m_2^2)}
\nonumber\\
&=& p^\mu p^\nu B_{21}(m_1^2,m_2^2,p^2)
+ g^{\mu\nu} B_{22}(m_1^2,m_2^2,p^2)\,. \nonumber
\ea
The two last integrals can be reduced to simpler integrals $A$ and $B$ via
\ba
B_1(m^2,m^2,p^2)& = &\frac{1}{2}B(m_1^2,m_2^2,p^2)\,,
\nonumber\\
B_{22}(m^2,m^2,p^2)& = & \frac{1}{2(d-1)}\Big[
 A(m^2)+\left(2 m^2-\frac{1}{2}p^2\right) B(m^2,m^2,p^2)\Big]\,,
\nonumber\\
B_{21}(m^2,m^2,p^2)& = &\frac{1}{p^2}\left[
  A(m^2)+m^2 B(m^2,m^2,p^2)-d B_{22}(m^2,m^2,p^2)\right]\,.
\ea
We quote here only the equal mass case results relevant for this paper.
The explicit expression for $B$ is
\ba
\label{resultintegrals}
B(m^2,m^2,p^2) &=& 
\frac{1}{16\pi^2}\lambda_0 +\overline B(m^2,p^2)+{\cal O}(\epsilon)\,,
\nonumber\\
\overline B(m^2,p^2) &=&\frac{1}{16\pi^2}
\left(-1-m^2\log\frac{m^2}{\mu^2}\right)
+\bar{J}(m_2,p^2)\,,
\nonumber\\
\bar{J}(m^2,p^2) &=&
-\frac{1}{16\pi^2}\int_0^1 dx
\ln\left(\frac{m^2-x(1-x)p^2}{m^2}\right)\,,
\ea
The function $\bar{J}(m^2,p^2)$ is
\ba
\bar{J}(m^2,p^2)&=&\left \{ \begin{array}{ll}
2+\sigma\ln\left(\frac{\sigma-1}{\sigma+1}\right),&p^2<0,\\
2-2\sqrt{\frac{4}{x}-1}\cdot\mbox{arccot}
\left(\sqrt{\frac{4}{x}-1}\right),&0\le p^2<4m^2,\\
2+\sigma\ln\left(\frac{1-\sigma}{1+\sigma}\right)+i\pi\sigma,&
p^2>4m^2,
\end{array} \right.
\nonumber\\
&&
\sigma(x)=\sqrt{1-\frac{4}{x}},\quad x={m^2\over p^2}\notin [0,4].
\ea
Taking derivatives w.r.t. $p^2$ at $p^2=0$
is most easily done in the form with the Feynman
parameter integration explicit.

\subsection{Sunset integrals}
\label{sunsetint}

The sunset integrals are done with the methods of \cite{ABT1,GS}.
They are defined as
\be
\lla X \rra
= \frac{1}{i^2}\int \frac{d^d q}{(2\pi)^d} \frac{d^d r}{(2\pi)^d}
\frac{X}
{\left(q^2-m_1^2\right)\left(r^2-m_2^2\right)\left[(q+r-p)^2-m_3^2\right]}\,,
\ee
The various sunset integrals with Lorenz indices are
\ba
\label{defsunset}
H(m_1^2,m_2^2,m_3^2;p^2) &=&\lla 1\rra\,,\nonumber\\
H^\mu(m_1^2,m_2^2,m_3^2;p^2) &=&\lla q^\mu\rra
= p^\mu H_1(m_1^2,m_2^2,m_3^2;p^2)\,,\\
H^{\mu\nu}(m_1^2,m_2^2,m_3^2;p^2) &=&\lla q^\mu q^\nu\rra\nonumber\\
& = &
p^\mu p^\nu H_{21}(m_1^2,m_2^2,m_3^2;p^2)
+g^{\mu\nu} H_{22}(m_1^2,m_2^2,m_3^2;p^2)
\,.
\nonumber
\ea
and
\ba
\lla r^\mu\rra &=& p^\mu H_1(m_2^2,m_1^2,m_3^2;p^2)\,,\nonumber\\
\lla r^\mu r^\nu\rra &=& p^\mu p^\nu H_{21}(m_2^2,m_1^2,m_3^2;p^2)
+ g^{\mu\nu} H_{22}(m_2^2,m_1^2,m_3^2;p^2)\,,\nonumber\\
\lla q^\mu r^\nu \rra &=& \lla r^\mu q^\nu\rra \,,\nonumber\\
\lla q^\mu r^\nu\rra &=&p^\mu p^\nu H_{23}(m_1^2,m_2^2,m_3^2;p^2)
+g^{\mu\nu} H_{24}(m_1^2,m_2^2,m_3^2;p^2)\,,
\ea
The function $H$ is fully symmetric in $m_1^2,m_2^2$ and $m_3^2$, while
$H_1$, $H_{21}$ and $H_{22}$ are symmetric under the interchange
of $m_2^2$ and $m_3^2$.
The relation between the above 3 functions
\ba
\label{H22relation}
\lefteqn{
p^2 H_{21}(m_1^2,m_2^2,m_3^2;p^2)+d H_{22}(m_1^2,m_2^2,m_3^2;p^2)
= }
&&\nonumber\\&&
 m_1^2 H(m_1^2,m_2^2,m_3^2;p^2)+A(m_2^2) A(m_3^2)\,,
\ea
allows to express $H_{22}$ in terms of $H_{21}$.

Similar to the integral $B$ and $B_1$, there is also a relation between $H$ and $H_1$ which in the equal mass case becomes
\be
\label{relationH1}
H_1(m^2,m^2,m^2;p^2) = \frac{1}{3}
 H(m^2,m^2,m^2;p^2)\,.
\ee
The other functions, $H_{23}$ and $H_{23}$, can be written in term of
$H$, $H_1$ and $H_{21}$ by using relations derived
from redefining the momenta and masses in its definition \cite{ABT1}.

The full sunset integral expressions and the definition
for finite part $H_i^F = \{H^F,H^F_1,H^F_{21}\}$ can be found in the appendix of \cite{ABT1}.
In our case we take $m_1 = m_2 = m_3 = m$.

In order to eliminate the extra poles in the expressions, sometimes we need to
expand the $H^F_i(m^2,m^2,m^2;q^2)$ around the pseudoscalar mass $m^2$,
and we define
\ba
H^M_i(m^2,m^2,m^2;q^2)&=&{1\over (q^2-m^2)^2}\Bigg[H^F_i(m^2,m^2,m^2;q^2)
-H^F_i(m^2,m^2,m^2;m^2)
\nonumber\\&&
-(q^2-m^2)H^{F\prime}_i(m^2,m^2,m^2;m^2)\Bigg]
\,,
\ea
where
\be
H^{F\prime}_i(m^2,m^2,m^2;m^2) = \frac{\partial H^F_i(m^2,m^2,m^2;q^2)}{\partial q^2}\Bigg|_{q^2=m^2}
\,.
\ee


\begin{thebibliography}{99}

\bibitem{Dimopoulos}
  S.~Dimopoulos,
  %``Technicolored Signatures,''
  Nucl.\ Phys.\  B {\bf 168} (1980) 69.

\bibitem{Peskin}
  M.~E.~Peskin,
  %``The Alignment Of The Vacuum In Theories Of Technicolor,''
  Nucl.\ Phys.\  B {\bf 175} (1980) 197.

\bibitem{Preskill}
  J.~Preskill,
  %``Subgroup Alignment In Hypercolor Theories,''
  Nucl.\ Phys.\  B {\bf 177}, 21 (1981).

\bibitem{Kogut}
  J.~B.~Kogut, M.~A.~Stephanov, D.~Toublan, J.~J.~M.~Verbaarschot and A.~Zhitnitsky,
  %``QCD-like theories at finite baryon density,''
  Nucl.\ Phys.\  B {\bf 582} (2000) 477
  [arXiv:hep-ph/0001171].

\bibitem{Kogan}
  Y.~I.~Kogan, M.~A.~Shifman and M.~I.~Vysotsky,
  %``Spontaneous Breaking Of Chiral Symmetry For Real Fermions And N=2 Susy
  %Yang-Mills Theory,''
  Sov.\ J.\ Nucl.\ Phys.\  {\bf 42} (1985) 318
  [Yad.\ Fiz.\  {\bf 42} (1985) 504].

\bibitem{Leutwyler}
  H.~Leutwyler and A.~V.~Smilga,
  %``Spectrum of Dirac operator and role of winding number in QCD,''
  Phys.\ Rev.\  D {\bf 46} (1992) 5607.

\bibitem{SV}
  A.~V.~Smilga and J.~J.~M.~Verbaarschot,
  %``Spectral Sum Rules And Finite Volume Partition Function In Gauge Theories
  %With Real And Pseudoreal Fermions,''
  Phys.\ Rev.\  D {\bf 51} (1995) 829
  [arXiv:hep-th/9404031].

\bibitem{GL2}
  J.~Gasser and H.~Leutwyler,
  %``Chiral Perturbation Theory: Expansions In The Mass Of The Strange Quark,''
  Nucl.\ Phys.\  B {\bf 250} (1985) 465.

\bibitem{GL4}
  J.~Gasser and H.~Leutwyler,
  %``Light Quarks at Low Temperatures,''
  Phys.\ Lett.\  B {\bf 184} (1987) 83.

\bibitem{Splittorff}
  K.~Splittorff, D.~Toublan and J.~J.~M.~Verbaarschot,
  %``Diquark condensate in QCD with two colors at next-to-leading order,''
  Nucl.\ Phys.\  B {\bf 620} (2002) 290
  [arXiv:hep-ph/0108040].

\bibitem{paper1}
  J.~Bijnens and J.~Lu,
  %``Technicolor and other QCD-like theories at next-to-next-to-leading order,''
  JHEP {\bf 0911} (2009) 116
  [arXiv:0910.5424 [hep-ph]].

\bibitem{paper2}
  J.~Bijnens and J.~Lu,
  %``Meson-meson Scattering in QCD-like Theories,''
  JHEP {\bf 1103} (2011) 028
  [arXiv:1102.0172[hep-ph]].

\bibitem{Techni1}
  J.~R.~Andersen, O.~Antipin, G.~Azuelos, L.~Del Debbio, E.~Del Nobile, S.~Di Chiara, T.~Hapola, M.~Jarvinen {\it et al.},
  %``Discovering Technicolor,''
  Eur.\ Phys.\ J.\ Plus {\bf 126 } (2011)  81.
  [arXiv:1104.1255 [hep-ph]].

\bibitem{Techni2}
  C.~T.~Hill and E.~H.~Simmons,
  %``Strong dynamics and electroweak symmetry breaking,''
  Phys.\ Rept.\  {\bf 381} (2003) 235
  [Erratum-ibid.\  {\bf 390} (2004) 553]
  [arXiv:hep-ph/0203079].

\bibitem{Peskin:1991sw}
  M.~E.~Peskin, T.~Takeuchi,
  %``Estimation of oblique electroweak corrections,''
  Phys.\ Rev.\  {\bf D46 } (1992)  381-409.

\bibitem{Altarelli:1990zd}
  G.~Altarelli, R.~Barbieri,
  %``Vacuum polarization effects of new physics on electroweak processes,''
  Phys.\ Lett.\  {\bf B253 } (1991)  161-167.

\bibitem{Weinberg1}
  S.~Weinberg,
  %``Phenomenological Lagrangians,''
  Physica A {\bf 96} (1979) 327.

\bibitem{GL1}
  J.~Gasser and H.~Leutwyler,
  %``Chiral Perturbation Theory To One Loop,''
  Annals Phys.\  {\bf 158} (1984) 142.

\bibitem{CCWZ}
  S.~R.~Coleman, J.~Wess and B.~Zumino,
  %``Structure of phenomenological Lagrangians. 1,''
  Phys.\ Rev.\  {\bf 177} (1969) 2239;
  C.~G.~.~Callan, S.~R.~Coleman, J.~Wess and B.~Zumino,
  %``Structure of phenomenological Lagrangians. 2,''
  Phys.\ Rev.\  {\bf 177} (1969) 2247.

\bibitem{BCE1}
  J.~Bijnens, G.~Colangelo and G.~Ecker,
  %``The mesonic chiral Lagrangian of order p**6,''
  JHEP {\bf 9902} (1999) 020
  [arXiv:hep-ph/9902437].

\bibitem{BCE2}
  J.~Bijnens, G.~Colangelo and G.~Ecker,
  %``Renormalization of chiral perturbation theory to order p**6,''
  Annals Phys.\  {\bf 280} (2000) 100
  [arXiv:hep-ph/9907333].

\bibitem{ABT1}
G.~Amoros, J.~Bijnens and P.~Talavera,
%``Two-point functions at two loops in three flavour chiral perturbation  theory,''
{\em Nucl. Phys.} B { 568} (2000) 319 [hep-ph/9907264].

\bibitem{GK1}
  E.~Golowich, J.~Kambor,
  %``Two loop analysis of vector current propagators in chiral perturbation theory,''
  Nucl.\ Phys.\  {\bf B447 } (1995)  373-404.
  [hep-ph/9501318].

\bibitem{GK}
E.~Golowich and J.~Kambor,
% ``Two-loop analysis of axialvector current propagators in chiral
% perturbation theory,''
{\em Phys. Rev.}\ D {58}, 036004 (1998) [hep-ph/9710214].

\bibitem{BT}
  J.~Bijnens, P.~Talavera,
  %``Pion and kaon electromagnetic form-factors,''
  JHEP {\bf 0203 } (2002)  046.
  [hep-ph/0203049].

\bibitem{GPP}
  M.~Gonzalez-Alonso, A.~Pich, J.~Prades,
  %``Determination of the Chiral Couplings L(10) and C(87) from Semileptonic Tau Decays,''
  Phys.\ Rev.\  {\bf D78 } (2008)  116012.
  [arXiv:0810.0760 [hep-ph]].

\bibitem{FORM}
  J.~A.~M.~Vermaseren,
  %``New features of FORM,''
  arXiv:math-ph/0010025.

\bibitem{GS}
  J.~Gasser and M.~E.~Sainio,
  %``Two-loop integrals in chiral perturbation theory,''
  Eur.\ Phys.\ J.\  C {\bf 6} (1999) 297
  [arXiv:hep-ph/9803251].

\end{thebibliography}
\end{document}